\documentclass[twocolumn]{aastex63}

\accepted{February 1, 2023}
\submitjournal{\apj}

\newcommand{\cha}{\textit{Chandra }}
\def\xmm{{XMM-{\it Newton\/}}}
\def\fermi{{{\it Fermi}-LAT }}

\def\apj{ApJ}
\def\aap{A\&A}

\def\apjl{ApJL}
\def\apjs{ApJS}

\shorttitle{Gamma-ray Emission from Pulsar Wind Nebula B0453-685 Discovered by the \fermi}

\shortauthors{Eagle et al.}

\PassOptionsToPackage{colorlinks,linkcolor={blue},citecolor={blue},urlcolor={red}}{hyperref}
\usepackage{hyperref}
\usepackage{natbib}
\usepackage{float}
\usepackage{color}
\usepackage{graphicx}
\usepackage{gensymb}
\usepackage{amsmath}
\usepackage{enumitem}
\usepackage{url}
\usepackage{multirow,helvet}
\usepackage{todonotes}
\usepackage{lineno}

\begin{document}



\title{Fermi--LAT Gamma-ray Emission Discovered from the Composite Supernova Remnant B0453--685 in the Large Magellanic Cloud}

\correspondingauthor{Jordan Eagle}
\email{jordan.l.eagle@nasa.gov}

\author[0000-0001-9633-3165]{Jordan Eagle}
\affiliation{Harvard-Smithsonian Center for Astrophysics\\
Cambridge, MA 02138, USA}
\affiliation{Department of Physics \& Astronomy\\ 
Clemson University\\ 
Clemson, SC 29634, USA}

\author[0000-0002-0394-3173]{Daniel Castro}
\affiliation{Harvard-Smithsonian Center for Astrophysics\\
Cambridge, MA 02138, USA}

\author{Peter Mahhov}
\affiliation{New York University Abu Dhabi, P.O. Box 129188, Abu Dhabi, United Arab Emirates\\}

\author[0000-0003-4679-1058]{Joseph Gelfand}
\affiliation{New York University Abu Dhabi, P.O. Box 129188, Abu Dhabi, United Arab Emirates\\}

\author[0000-0002-0893-4073]{Matthew Kerr}
\affiliation{Space Science Division, Naval Research Laboratory, Washington, DC 20375-5352, USA\\ }

\author[0000-0002-6986-6756]{Patrick Slane}
\affiliation{Harvard-Smithsonian Center for Astrophysics\\
Cambridge, MA 02138, USA}

\author[0000-0002-8784-2977]{Jean Ballet}
\affiliation{Universit\'{e} Paris-Saclay, Universit\'{e} Paris Cit\'{e}, CEA, CNRS, AIM, 91191 Gif-sur-Yvette, France \\}

\author[0000-0002-6606-2816]{Fabio Acero}
\affiliation{Universit\'{e} Paris-Saclay, Universit\'{e} Paris Cit\'{e}, CEA, CNRS, AIM, 91191 Gif-sur-Yvette, France \\}

\author[0000-0003-4136-7848]{Samayra Straal}
\affiliation{New York University Abu Dhabi, P.O. Box 129188, Abu Dhabi, United Arab Emirates\\}

\author[0000-0002-6584-1703]{Marco Ajello}
\affiliation{Department of Physics \& Astronomy\\ 
Clemson University\\ 
Clemson, SC 29634, USA}



\begin{abstract}
We report the second extragalactic pulsar wind nebula (PWN) to be detected in the MeV--GeV band by the {\it Fermi}--LAT, located within the Large Magellanic Cloud (LMC). 
The only other known PWN to emit in the Fermi band outside of the Milky Way Galaxy is N~157B which lies to the west of the newly detected $\gamma$-ray emission at an angular distance of 4\,$\degree$. Faint, point-like $\gamma$-ray emission is discovered at the location of the composite supernova remnant (SNR) B0453--685 with a $\sim$ 4\,$\sigma$ significance from energies 300\,MeV--2\,TeV. We present the \fermi data analysis of the new $\gamma$-ray source, coupled with a detailed multi-wavelength investigation to understand the nature of the observed emission. Combining the observed characteristics of the SNR and the physical implications from broadband modeling, we argue it is unlikely the SNR is responsible for the $\gamma$-ray emission. While the $\gamma$-ray emission is too faint for a pulsation search, we try to distinguish between any pulsar and PWN component of SNR~B0453--685 that would be responsible for the observed $\gamma$-ray emission using semi-analytic models. We determine the most likely scenario is that the old PWN ($\tau\sim 14,000$\,years) within B0453--685 has been impacted by the return of the SNR reverse shock with a possible substantial pulsar component below $5\,$GeV. 
\end{abstract}

\section{Introduction}
Pulsar wind nebulae (PWNe) are descendants of core collapse supernovae (CC SNe), each powered by an energetic, rapidly rotating neutron star. As the neutron star spins down, rotational energy is translated into a relativistic particle wind, made up of mostly electrons and positrons \citep{slane2017}. 
The evolution of a PWN is connected to the evolution of the central pulsar, host supernova remnant (SNR), and the structure of the surrounding interstellar medium \citep[ISM,][]{gaensler2006}. Eventually, the relativistic particle population will be injected into the ISM of the host galaxy and may contribute to the cosmic ray (CR) electron--positron population \citep{crflux}. 

Synchrotron emission from relativistic electrons is observed from the majority of PWNe, from radio wavelengths to hard X-rays. Moreover, CR electrons are expected to scatter off of local photon fields, resulting in Inverse Compton (IC) emission at $\gamma$-ray energies \citep{gaensler2006}. Accordingly, the majority of PWNe have been discovered in the radio or X-ray bands and an increasing number of discoveries are occurring in TeV $\gamma$-rays. In fact, the majority of the Galactic TeV source population is found to be PWNe as observed by Cherenkov Telescopes \citep[$\sim 37$, e.g.][]{tevcat2008,acero2013}. On the other hand, only 11 PWNe have been firmly identified in the MeV--GeV band with the \fermi \citep{atwood2009}. However, upgrades in the event processing of the \fermi data have significantly improved the spatial resolution and sensitivity of the instrument \citep[Pass 8,][]{atwood2013}.
Taking advantage of the upgrade and using \fermi observations with $\sim$ 11.5\,years of data, we have discovered a new \fermi $\gamma$-ray source located in the Large Magellanic Cloud (LMC) and belongs to the composite SNR~B0453--685. We combine the new $\gamma$-ray measurements with available multi-wavelength data for the region to determine that the PWN is the most likely origin of the $\gamma$-rays and that a pulsar may contribute to the lower-energy $\gamma$-ray emission.

The broadband spectrum of a PWN depends both on the particle spectrum that was initially injected by the pulsar and how it was altered throughout the evolution of the PWN inside its SNR \citep{reynolds_1984,gelfand_2009}. In order to rigorously explore the characteristics of the underlying particle population(s), we present a semi-analytic simulation for the dynamical and radiative evolution of a PWN inside an SNR. 

In Section \ref{sec:select} we describe the SNR~B0453--685 system. We present a multi-wavelength analysis in Section \ref{sec:mw}, describing the X-ray analysis using archival \cha observations in Section \ref{sec:chamethod} and the \fermi data analysis in Section \ref{sec:fermethod}. We present simple broadband models investigating the $\gamma$-ray origin in Section~\ref{sec:naima_model}. We further simulate a broadband spectral model using a semi-analytic model for PWN evolution, which incorporates known properties of the system and report the resulting best-fit spectral energy distribution (SED) in Section \ref{sec:yosi_model}. We discuss implications of observations and modeling and we provide our final conclusions in Sections \ref{sec:discuss} and \ref{sec:conclude}.

\begin{figure*}
\begin{minipage}[b]{.33\textwidth}
\hspace{-1.33cm}
\vspace{-0.55cm}
\includegraphics[width=1.18\linewidth]{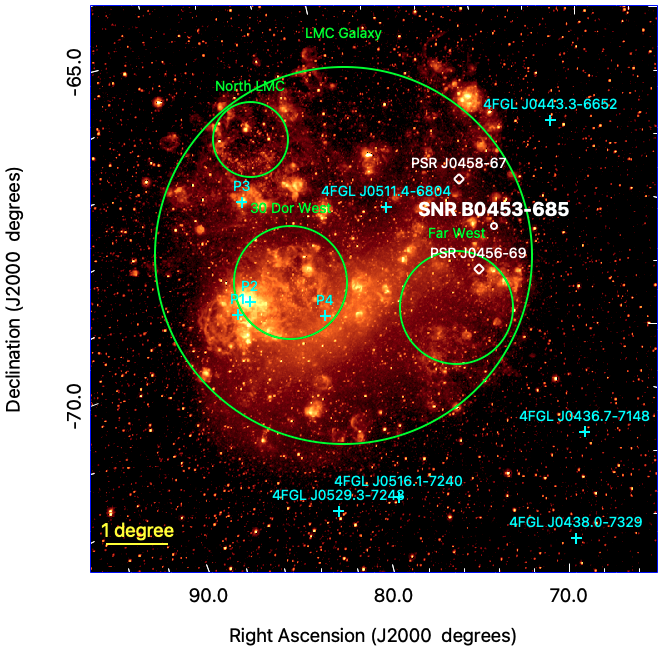}
\end{minipage}
\begin{minipage}[b]{.33\textwidth}
\hspace{-0.25cm}
\includegraphics[width=1.07\linewidth]{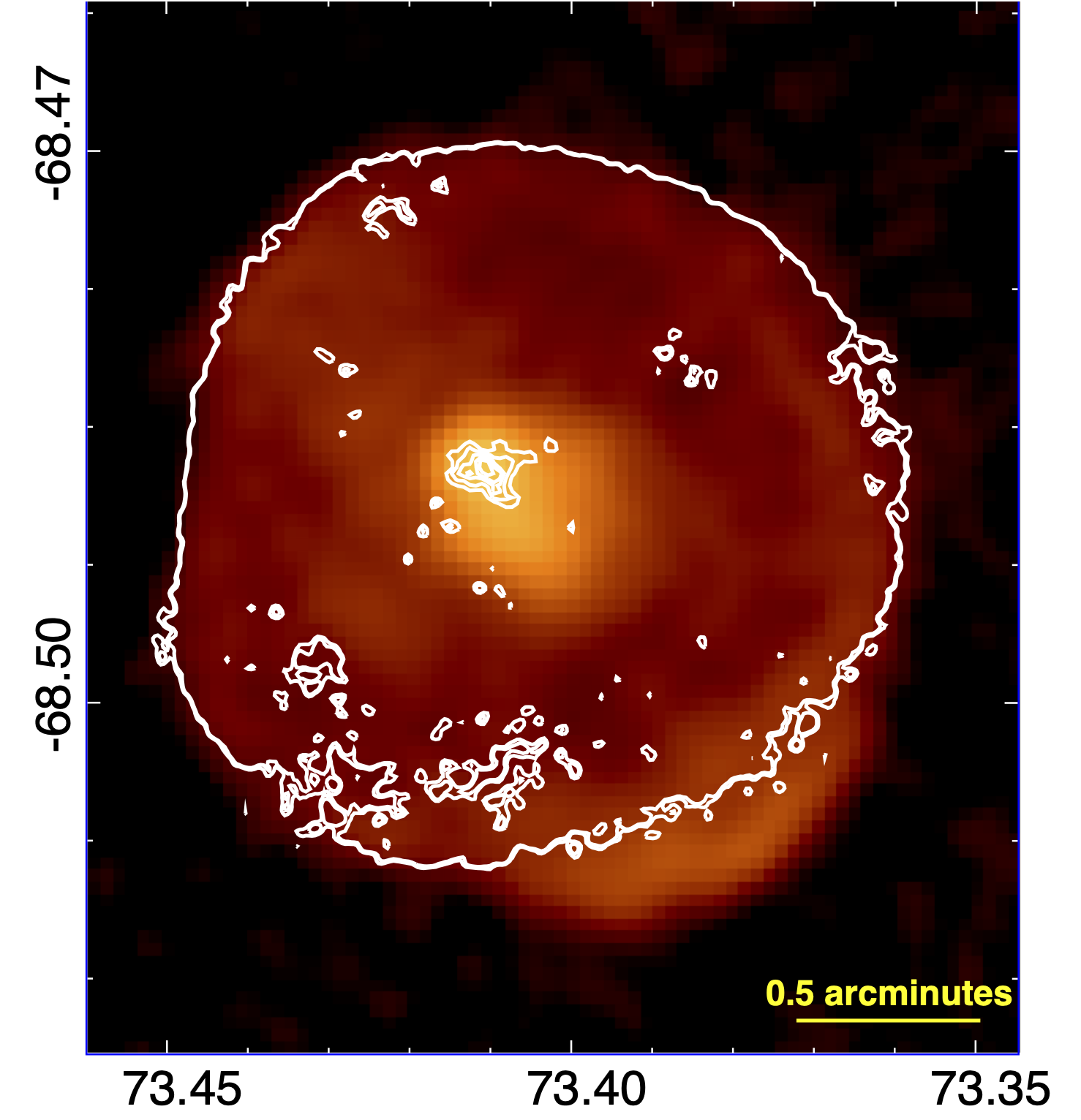}
\end{minipage}
\begin{minipage}[b]{.33\textwidth}
\hspace{-0.15cm}
\includegraphics[width=1.2\linewidth]{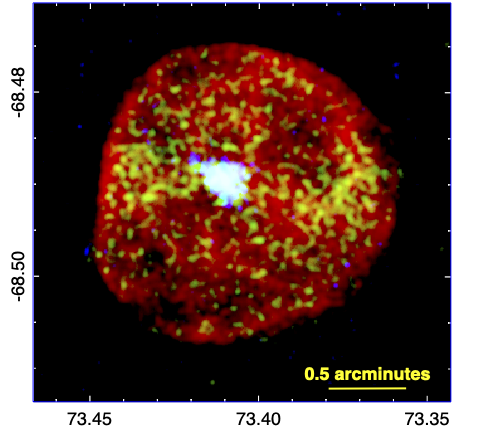}
\end{minipage}
\caption{{\it Left:} The SAO DS9 image of the LMC in the H$\alpha$ band from the Southern H-Alpha Sky Survey Atlas$^a$ \citep[SHASSA,][]{halpha2001}. The P1--P4 labels identify the four brightest 4FGL point sources in the LMC, following the naming convention used in \citet{lmc2016}. P1 is the most energetic pulsar ever detected, PSR~J0540-6919, which lies $< 0.5\,\degree$ from P2. P2 is the possible \fermi PWN N~157B. P3 is a high mass binary (HMB) system and P4 is the SNR N~132D located near the 30 Doradus region. The four extended templates used to describe the diffuse $\gamma$-ray emission from the LMC \citep[components E1--E4 in][]{lmc2016} are indicated with the green circles. The location of SNR~B0453--685 is marked in white with radius $r=0.05\degree$. The two closest known radio pulsars near SNR~B0453--685 are labeled as white diamonds$^b$. Both are located too far from the SNR to be a reasonable central pulsar candidate. The coordinates are labeled and are in equatorial J2000 degrees throughout the paper unless otherwise noted. {\it Middle:} The SAO DS9 image of the 1.4\,GHz radio emission observed from SNR~B0453--685 \citep{gaensler2003}. The white contours correspond to the central PWN and the outer SNR shell as observed in X-ray (right panel). {\it Right:} Tri-color X-ray flux map generated in SAO DS9 of SNR~B0453--685 \citep{gaensler2003}. Red is soft X-ray emission between 0.5--1.2\,keV, green is medium flux between 1.2--2\,keV, and blue is hard flux from 2--8\,keV. Soft and medium X-ray emission outlines and fills the entire SNR while the hard X-ray emission is heavily concentrated towards the center of the SNR where the PWN is located.\footnotesize{$^a$ SHASSA is supported by the National Science Foundation. $^b$ We used the ATNF radio pulsar catalog \url{https://www.atnf.csiro.au/research/pulsar/psrcat/} \citep{manchester2005}.}}\label{fig:radio_xray_halpha}
\end{figure*}

\section{SNR~B0453--685}\label{sec:select}
SNR~B0453--685 is located in the LMC with a distance $d\approx50$\,kpc \citep{clementini2003}. The LMC has an angular size of nearly 6 degrees in the sky where SNR~B0453--685 (angular size $r <$ 0.05\,$\degree$) is positioned on the western wall of H$\alpha$ emission as shown in Figure~\ref{fig:radio_xray_halpha}, left panel. SNR~B0453-685 was identified as a middle-aged ($\tau \sim 13\,$kyr) composite SNR hosting a bright, polarized central core by \citet{gaensler2003} based on observations at 1.4 and 2.4\,GHz frequencies and in X-rays between 0.3--8.0\,keV; see the middle and right panels of Figure~\ref{fig:radio_xray_halpha}. A thin, faint SNR shell is visible in both radio and X-ray (0.3--2.0\,keV) with the softer, diffuse X-ray emission filling the SNR. Within the radio SNR shell is a much brighter, large, and polarized, central core: the PWN. The PWN also dominates the hard X-ray emission (2.0--8.0\,keV, Figure~\ref{fig:radio_xray_halpha}). While the radio and X-ray observations reported by \citet{gaensler2003} indicated the composite morphology of the SNR, no pulsations from a central pulsar have been detected. 

\citet{manchester2006} performed a deep radio pulsar search in both of the Magellanic Clouds with the Parkes 64--m radio telescope and reported 14 total pulsars, 11 of which were located within the LMC, but none were associated to SNR~B0453--685. It is reported in later investigations \citep[e.g.,][]{lopez2011,mcentaffer2012} using the same \cha X-ray observations as those in \citet{gaensler2003} that an X-ray point source is detected inside the central PWN core using the \texttt{wavdetect} tool within the \cha data reduction software package, CIAO \citep{ciao2006}. This remains the most promising evidence for the central pulsar. 

Displayed in Figure \ref{fig:radio_xray_halpha}, left panel, are the few known sources within the LMC that emit $\gamma$-rays in the \fermi band, labeled P1--P4 following the convention of \citet{lmc2016}. Only one LMC PWN, N~157B (P2), is identified as a GeV \citep{4fgl-dr2} and TeV \citep{hess2012} $\gamma$-ray source and it is located on the opposite (Eastern) wall of the LMC with respect to SNR~B0453--685. N~157B is located in a very crowded area, accompanied by two bright $\gamma$-ray sources nearby, SNR~N132~D and PSR~J0540--6919. SNR~B0453--685, however, is conveniently located in a much less crowded region of the LMC, making its faint point-like $\gamma$-ray emission detectable even against the diffuse LMC background, diffuse Galactic foreground, and the isotropic background emissions. 

\section{Multiwavelength Information}\label{sec:mw}

\subsection{Radio}
Australia Telescope Compact Array (ATCA) observations at 1.4 and 2.4\,GHz 
 revealed the composite nature of SNR~B0453--685, indicating the presence of a PWN \citep{gaensler2003}. The PWN is visible as a bright central core that is surrounded by the SNR shell roughly 2$^{\prime}$ in diameter. \citet{gaensler2003} measure the flux density of the radio core to be 46$ \pm 2$\,mJy at both 1.4 and 2.4\,GHz. The PWN radio spectrum is flat, with $\alpha = -0.10\pm0.05$ \citep{gaensler2003}. No central point source such as a pulsar is seen, but the authors place an upper limit on a point source of 3\,mJy at 1.4\,GHz and 0.4\,mJy at 2.4\,GHz at the location of the emission peak and suggest the PWN to be powered by a Vela-like pulsar with a spin period of $P \approx 100$\,ms, a surface magnetic field $B \approx 3\times10^{12}$\,Gauss, and a spin-down luminosity $\dot{E} \approx 10^{37}$\, ergs s$^{-1}$. 

\citet{haberl2012} observed SNR~B0453--685 with ATCA at 4.8 and 8.6 \,GHz, providing radio flux density measurements of both the SNR and PWN. The authors measure a flat radio spectrum for the PWN, with $\alpha_{pwn} =- 0.04 \pm 0.04$, along with significant polarization from the PWN core at 1.4\,GHz, 2.4\,GHz, 4.8\,GHz, and 8.6\,GHz frequencies. The outer SNR shell, excluding the PWN contribution, has a radio spectral index $\alpha_{shell} = -0.43 \pm 0.01$, which is a typical value for radio SNR shells.

\subsection{X-ray}\label{sec:chamethod}

\subsubsection{\cha X-ray Data Analysis}
\begin{figure}
\centering
\includegraphics[width=1.0\linewidth]{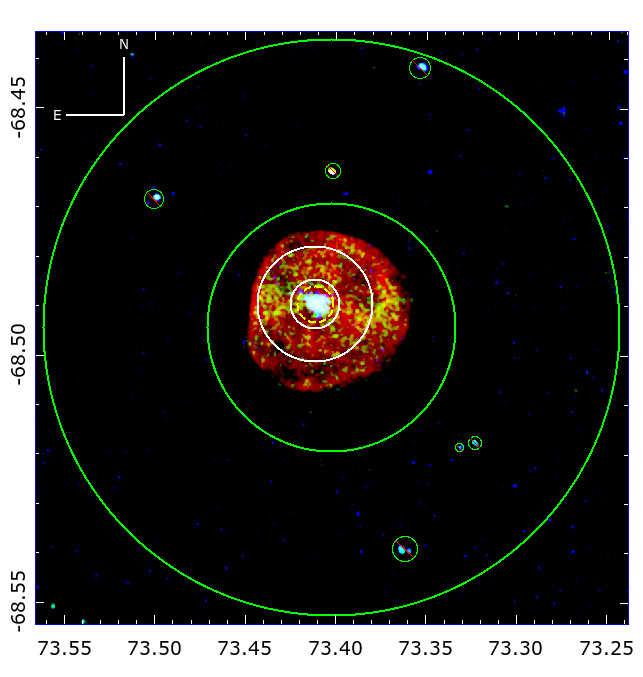}
\caption{Tri-color X-ray flux map generated in SAO DS9 of B0453--685. Red = 0.5--1.2\,keV, green is 1.2--2\,keV and blue is 2--8\,keV. The source and background regions used for spectral analysis are indicated. The yellow dashed circle corresponds to the PWN region, the white annulus corresponds to the SNR region, and the large green annulus excluding six bright X-ray point sources corresponds to the background region.}\label{fig:chandra}
\end{figure}

SNR~B0453--685 has been analyzed in X-rays in great detail \citep{gaensler2003, lopez2009, haberl2012, mcentaffer2012} with data from \xmm\ and \cha X-ray telescopes. Thermal X-ray emission dominates the soft X-rays and is largely attributed to the SNR while the hard X-ray emission is concentrated towards the center of the remnant where the PWN is located (see Figure \ref{fig:radio_xray_halpha}, right). In order to understand the $\gamma$-ray origin, we must combine the new \fermi data with available multi-wavelength data for the region. Therefore, we re-analyzed archival Chandra X-ray observations (ObsID: 1990) taken with the Advanced CCD Imaging Spectrometer (ACIS) on board the Chandra X-ray Observatory. The observation exposure is 40\,ks and was completed on 2001 December 18. The entire SNR is imaged on one back-illuminated chip (called ``S3'', see Figure \ref{fig:chandra}). 
Data reprocessing was conducted using the standard processing procedures in the Chandra Interactive Analysis of Observations \citep[CIAO v.4.12,][]{ciao2006} software package. The cleaned spectra are then extracted and background-subtracted using one large annulus-shaped region surrounding the remnant. We model both SNR and PWN emission components using data extracted from the regions indicated in Figure \ref{fig:chandra} and perform a spectral analysis. A spectrum for each component is extracted using the \texttt{specextract} tool in CIAO and modeled using SHERPA within CIAO \citep{sherpa2001}.

\subsubsection{\cha X-ray Data Analysis Results}\label{sec:cha-results}

\begin{figure}
\hspace{-0.5cm}
\includegraphics[width=1.05\linewidth]{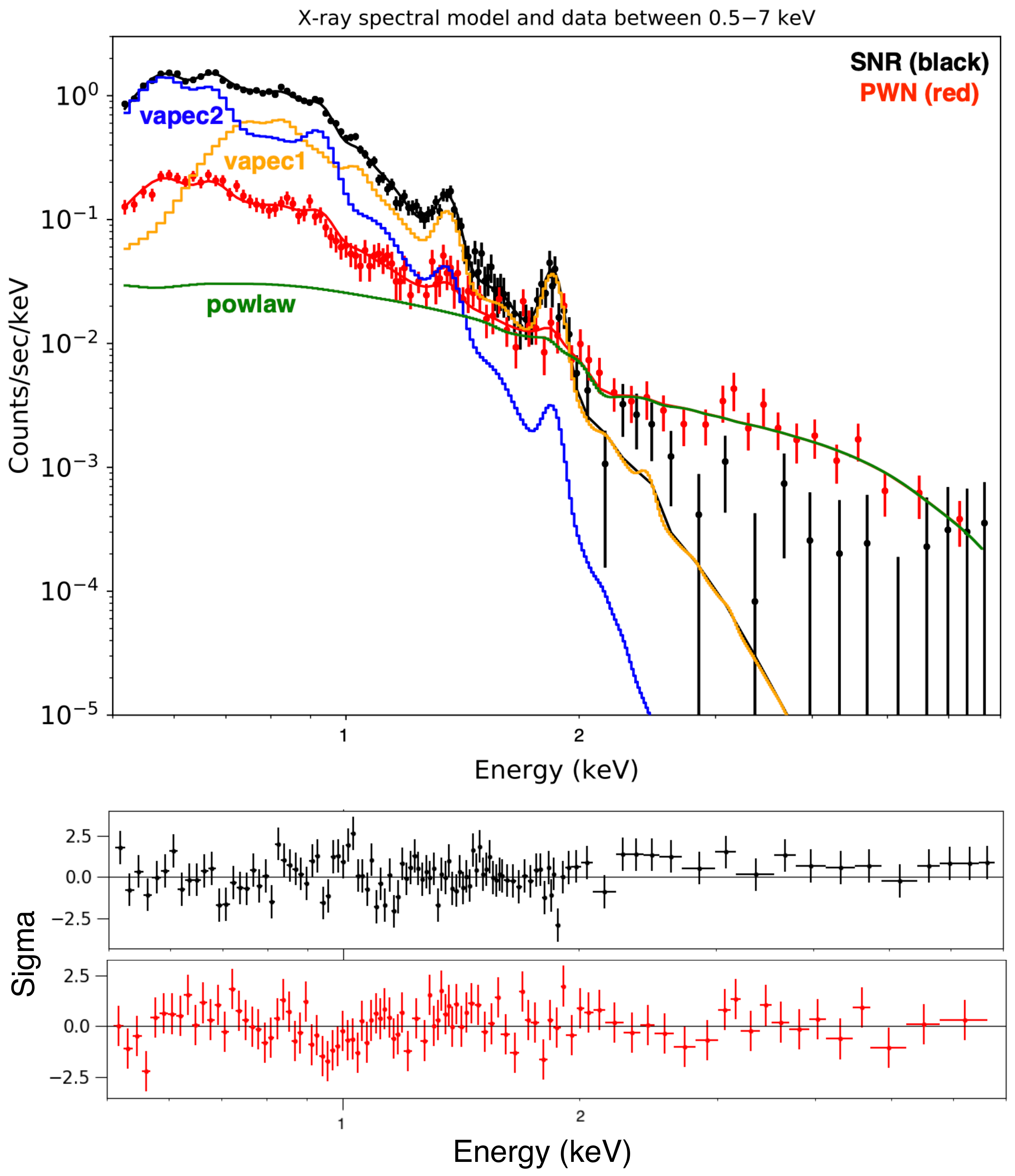}
\caption{{\it Top:} 0.5--7\,keV X-ray data and best-fit models for the two source models. The green solid line represents the non-thermal component from the PWN and the solid orange and blue lines represent the first and second thermal components of the SNR spectrum, respectively. {\it Bottom:} The residuals of the difference in the best-fit model and data for the SNR spectral fit (black,top) and the PWN spectral fit (red,bottom) in units of $\sigma$.}\label{fig:xray_models}
\end{figure}

Data between 0.5--7\,keV are used to model observed emission and is binned to at least 20 counts per bin. We fit the two source regions for the SNR and PWN components simultaneously and the best-fit model is displayed in Figure \ref{fig:xray_models}. A two-component collisionally ionized plasma model (\texttt{xsvapec}) is found to best describe the emission from the SNR and one nonthermal \texttt{powlaw1d} model is preferred for the PWN component \citep[similar to prior works, e.g.][]{haberl2012, mcentaffer2012}. We account for interstellar absorption along the line of sight by including the \texttt{tbabs} hydrogen column density parameter which uses the abundances estimated from \citet{wilms2000}. The best-fit parameters are listed in Table \ref{tab:xrayparams} along with the corresponding 90\% confidence intervals using the \texttt{conf} tool in Sherpa. 

The initial values of elemental abundances are set to those estimated for the LMC in \citet{lmc1992} and are allowed to vary one by one in each fit iteration. We keep the abundance of an element free if it significantly improves the fit, otherwise the value remains frozen at the following abundances relative to solar: He 0.89, C 0.26, N 0.16, O 0.32, Ne 0.42, Mg 0.74, Si 1.7, S 0.27, Ar 0.49, Ca 0.33, Fe 0.50, and Ni 0.62. Aluminum is not well constrained \citep[see Section 4.3 in][]{lmc1992} so we freeze its value to 0.33. 

\begin{table*}[t!]
\centering
\begin{tabular}{|c c c|}
\hline
\ Data points & DOF$^a$ & Reduced $\chi^2$ \\
\hline 
204 & 191 & 0.94 \\
\hline
\hline
\ Component & Model & \\
\hline
\ SNR & {\texttt{tbabs$\times$(vapec$_1$+vapec$_2$)}} & \\
\ PWN & {\texttt{tbabs$\times$\big[(c$_1 \times$vapec$_1$+c$_2 \times$vapec$_2$) + powlaw\big]}} & \\
\hline
\end{tabular}
\caption{Summary of the statistics and best-fit model for the SNR and PWN components in the X-ray analysis. The thermal components of the PWN spectrum is linked to the SNR model with the free coefficients $c_1$ and $c_2$. \footnotesize{$^a$ Degrees of freedom}}
\label{tab:xraymodel}
\end{table*}

The PWN spectrum is contaminated by two thermal components from the SNR emission in addition to a nonthermal component described best as a power law. Because SNR emission contaminates the PWN emission, we link the thermal parameters of the two models using the \texttt{scale1d} parameter in Sherpa (Table \ref{tab:xraymodel}). We leave the amplitude, $C_0$, free to vary in the fit for both thermal components.

\begin{table*}
\centering
\begin{tabular}{|c c c |}
\hline
\ SNR & & \\
\ Component & Parameter & Best-Fit Value \\
\hline
\ {\texttt{tbabs}}$^a$ & N$_{H}$(10$^{22}$ cm$^{-2}$) & 0.37$_{-0.09}^{+0.11}$ \\
\hline
\ \texttt{vapec$_1$} & $kT$(keV) & 0.34$_{-0.05}^{+0.02}$ \\
\ & Normalization & 3.67$_{-0.97}^{+2.55}\times10^{-3}$ \\
\ \texttt{vapec$_2$} & $kT$(keV) & 0.16$_{-0.01}^{+0.01}$ \\
\ & O & 0.35$_{-0.11}^{+0.26}$ \\
\ & Ne & 0.39$_{-0.13}^{+0.32}$ \\
\ & Mg & 0.56$_{-0.33}^{+0.50}$ \\
\ & Fe & $<$ 0.70 \\
\ & Normalization & 0.05$_{-0.03}^{+0.06}$ \\
\hline
\hline
\ PWN & & \\
\ Component & Parameter & Best-Fit Value \\
\hline
\ $c_1$ & $C_0$ & 0.07$_{-0.05}^{+0.02}$ \\
\ $c_2$ & $C_0$ & 0.14$_{-0.01}^{+0.01}$ \\
\ \texttt{powlaw} & $\Gamma$ & 1.74$_{-0.20}^{+0.20}$ \\
\ & Amplitude & 5.28$_{-1.01}^{+1.18}\times10^{-5}$\\
\hline
\end{tabular}
\caption{Summary of the 90\% C.L. statistics and parameters for the best-fit model for each component in the X-ray analysis. Metal abundances are reported in solar units. \footnotesize{$^a$ Absorption cross section set to \citet{vern1996}.}}
\label{tab:xrayparams}
\end{table*}

The hydrogen column density is $N_H = 3.7^{+1.1}_{-0.9} \times 10^{21}$\,cm$^{-2}$, the PWN power law index is $\Gamma_X = 1.74^{+0.20}_{-0.20}$, and the unabsorbed X-ray flux of the PWN component between 0.5--7\,keV is $f_x = 2.68 \pm 0.59 \times10^{-13}$\,erg cm$^{-2}$ s$^{-1}$. The $N_H$ value is reasonable compared to what is measured in the direction of the LMC\footnote{Using the \texttt{nh} tool from the HEASARC FTOOLS package \url{http://heasarc.gsfc.nasa.gov/ftools}.}, $N_H = 2.2\times10^{21}$\,cm$^{-2}$ \citep{ftools1995}. The best-fit model is consistent with other X-ray analyses \citep{haberl2012, mcentaffer2012}, with the largest differences being the elemental abundances, which can be explained by the use of \citet{wilms2000} and the \citet{vern1996} cross sections in this work, in addition to slight differences in choice of model components for the thermal emission and detector capabilities. In particular, \citet{haberl2012} analyzed \xmm\ observations of the entire SNR, but the PWN is not resolved and thus only one global spectrum was used to characterize any SNR and PWN emission. The SNR is much brighter than the PWN in X-rays so the nonthermal component from the PWN in the \xmm\ X-ray spectrum is not well constrained. \citet{mcentaffer2012} used \citet{anders} abundances and \citet{bal1992} cross-sections, and instead of two thermal equilibrium models, \texttt{vapec}, their best-fit model assumes a two-component structure from a \texttt{vapec+vnei} combination, where the \texttt{vnei} models the second thermal component without ionization equilibrium conditions. 

The best-fit temperatures for the two-component thermal model used to describe SNR emission are $kT = 0.34^{+0.02}_{-0.05}$\,keV and $kT = 0.16^{+0.01}_{-0.01}$\,keV, similar to what is reported in \citet{mcentaffer2012}. 
The PWN spectrum is non-thermal and best fit with a power law and photon index, $\Gamma_X = 1.74^{+0.20}_{-0.20}$. The PWN's spectral index is slightly harder than what is reported in \citet{mcentaffer2012}, where an index $\Gamma_X \sim 2$ across the PWN region is measured, but is still in agreement within the 90\% C.L. uncertainties. 
No synchrotron component is attributed to the SNR, but we estimate the 0.5--7\,keV 90\% C.L. upper limit of the flux for a nonthermal component to the SNR spectrum to be $F_X < 5.5 \times 10^{-13}$\,erg cm$^{-2}$ s$^{-1}$.


\subsection{Gamma-ray}\label{sec:fermethod}

\subsubsection{\fermi}
The {\it Fermi} Gamma-ray Space Telescope houses 
the Large Area Telescope \citep[LAT,][]{atwood2009}. The LAT instrument is sensitive to $\gamma$-rays with energies from 50\,MeV to $> 300$\,GeV \citep{4fgl2020} and has been continuously surveying the entire sky every 3\,hours since beginning operation in 2008 August. 

We use 11.5\, years (from 2008 August to 2020 January) of Pass~8 \texttt{SOURCE} class data \citep{atwood2013,pass82018} between 300\,MeV and 2\,TeV. Photons detected at zenith angles larger than 100\,$\degree$ were excluded to limit the contamination from $\gamma$-rays generated by cosmic ray (CR) interactions in the upper layers of Earth's atmosphere. 



\subsubsection{\fermi Data Analysis}

We perform a binned likelihood analysis with the latest Fermitools package\footnote{\url{https://fermi.gsfc.nasa.gov/ssc/data/analysis/software/}} (v.2.0.8) and FermiPy Python~3 package \citep[v.1.0.1,][]{fermipy2017}, utilizing the \texttt{P8R3\_SOURCE\_V3} instrument response function (IRF) and account for energy dispersion, to perform data reduction and analysis. 
We organize the events by PSF type using \texttt{evtype=4,8,16,32} to represent \texttt{PSF0, PSF1, PSF2}, and \texttt{PSF3} components. A binned likelihood analysis is performed on each event type and then combined into a global likelihood function for the region of interest (ROI) to represent all events\footnote{See FermiPy documentation for details: \url{https://fermipy.readthedocs.io/en/0.6.8/config.html}}. We fit the square 10$\degree$ ROI centered on the PWN position in equatorial coordinates using a pixel bin size $0.05\,\degree$ and 10 bins per decade in energy (38 total bins). The $\gamma$-ray sky for the ROI is modeled from the latest comprehensive \fermi source catalog based on 10\,years of data, 4FGL \citep[data release 2 (DR2),][]{4fgl-dr2} for point and extended sources\footnote{{\url{https://fermi.gsfc.nasa.gov/ssc/data/access/lat/10yr_catalog/}.}} that are within 15$\degree$ of the ROI center, as well as the latest Galactic diffuse and isotropic diffuse templates (\texttt{gll\_iem\_v07.fits} and \texttt{iso\_P8R3\_SOURCE\_V3\_v1.txt}, respectively)\footnote{LAT background models and appropriate instrument response functions: \url{https://fermi.gsfc.nasa.gov/ssc/data/access/lat/BackgroundModels.html}.}.

Because B0453--685 is located in the LMC, we need to properly account for 
 the diffuse emission from the LMC. We employ in the 4FGL source model four additional extended source components to reconstruct the {\it{emissivity model}} developed in \citet{lmc2016} to represent the diffuse LMC emission. The four additional sources are 4FGL~J0500.9--6945e (LMC Far West), 4FGL~J0519.9--6845e (LMC Galaxy), 4FGL~J0530.0-6900e (30 Dor West), and 4FGL~J0531.8--6639e (LMC North). These four extended templates along with the isotropic and Galactic diffuse templates define the total background for the ROI.

With the source model described above, we allow the background components and sources with test statistic (TS) $\geq 25$ and distances from the ROI center $\leq3.0$\,$\degree$ to vary in spectral index and normalization. We computed a series of diagnostic TS and count maps in order to search for and understand any residual $\gamma$-ray emission. The TS value is defined to be the natural logarithm of the ratio of the likelihood of one hypothesis (e.g. presence of one additional source) and the likelihood for the null hypothesis (e.g. absence of source):
\begin{equation}
  TS = 2 \times \log\big({\frac{{\mathcal{L}_{1}}}{{\mathcal{L}_{0}}}}\big)
\end{equation}
The TS value quantifies the significance for a source detection with a given set of location and spectral parameters and the significance of such a detection can be estimated by taking the square root of the TS value for 1 DOF \citep{mattox1996}. TS values $>25$ correspond to a detection significance $> 4 \sigma$ for 4 DOF. 

We generated the count and TS maps for the following energy ranges: 300\,MeV--2\,TeV, 1--10\,GeV, 10--100\,GeV, and 100\,GeV--2\,TeV. The motivation for increasing energy cuts stems from the improving PSF of the \fermi instrument with increasing energies\footnote{See \url{https://www.slac.stanford.edu/exp/glast/groups/canda/lat_Performance.htm} for a review on the dependence of PSF with energy for Pass 8 data.}. We inspected the TS maps for additional sources, finding a faint point-like $\gamma$-ray source coincident in location with B0453--685 and no known 4FGL counterpart\footnote{The closest 4FGL source is the probable unclassified blazar 4FGL J0511.4--6804 $\sim2$\,$\degree$ away.}.

\begin{figure*}
\begin{minipage}[b]{.5\textwidth}
\centering
\includegraphics[width=1.0\linewidth]{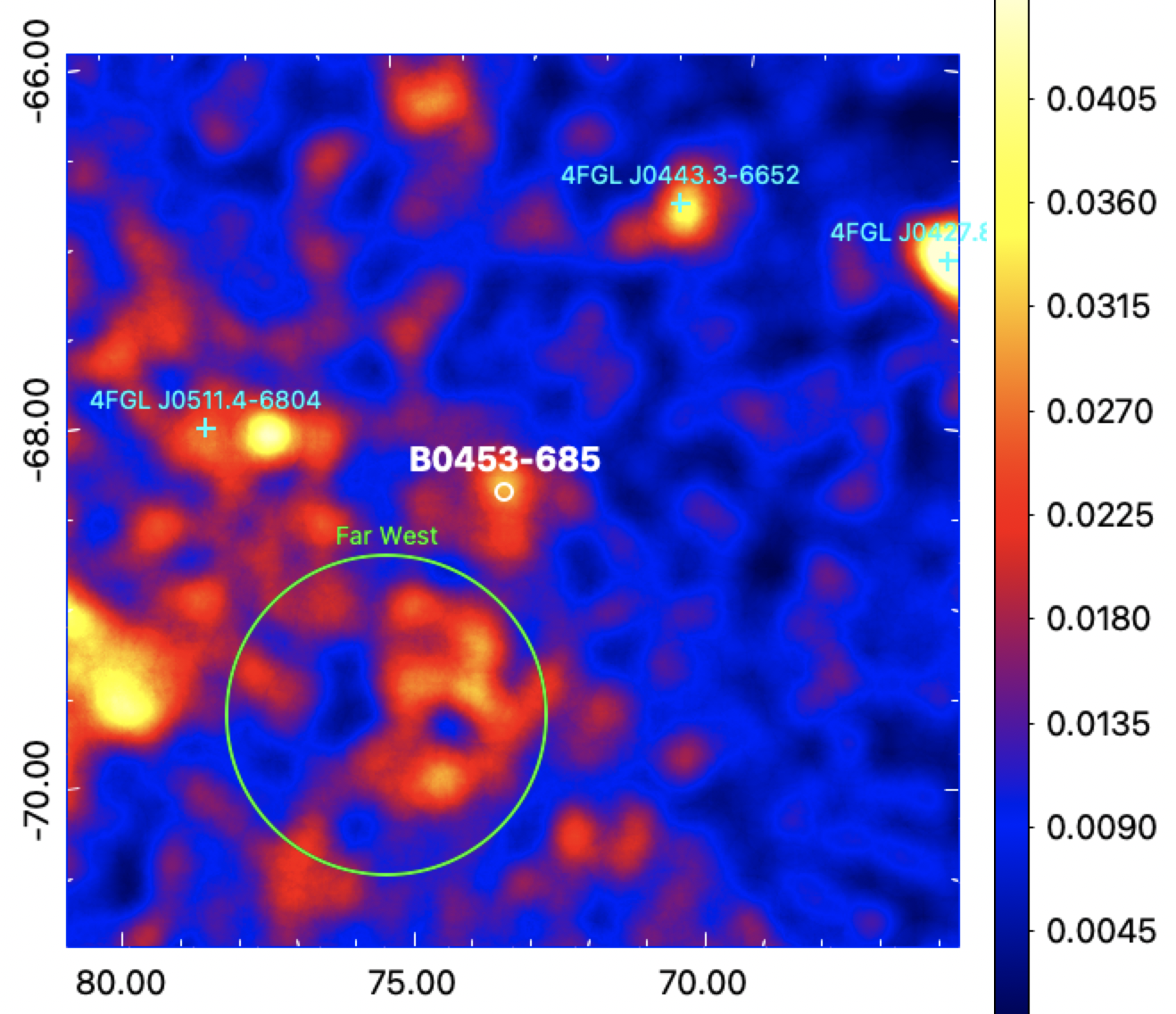}
\end{minipage}
\begin{minipage}[b]{.5\textwidth}
\hspace{-.4cm}
\centering
\includegraphics[width=1.0\linewidth]{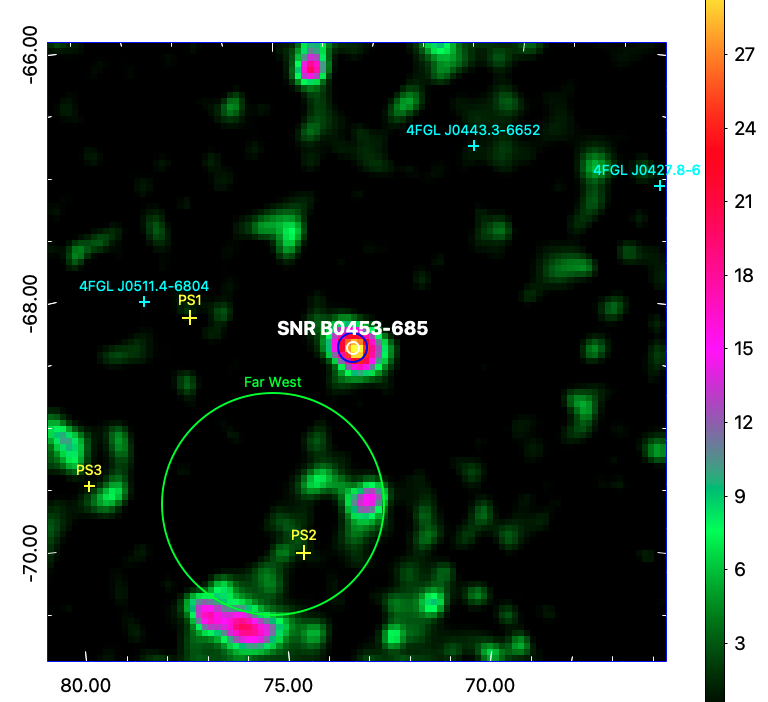}
\end{minipage}
\caption{{\it Left:} Smoothed ($\sigma = 0.1\,\degree$) 5\,$\degree\times 5$\,$\degree$ count map of \texttt{PSF3} events between 1--10\,GeV 
 with the locations of 4FGL sources in the field of view labeled. The pixel size is 0.01 deg pixel$^{-1}$. {\it Right:} 5\,$\degree\times 5$\,$\degree$ TS map between 1--10\,GeV. 
The maximum TS value at the SNR position is $\sim$ 28. The 95\% positional uncertainty for the best-fit $\gamma$-ray point source is in blue. 
In both panels, the location and approximate size of the composite SNR~B0453--685 ($r=0.02\degree$) is marked in white with radius $r=0.05\degree$.}\label{fig:gamma_maps}
\end{figure*}

Figure~\ref{fig:radio_xray_halpha}, left panel, demonstrates the total source model used in the analysis (except the isotropic and Galactic diffuse templates).  Three additional point sources are added to the source model that model residual emission in the field of view (PS1, PS2, and PS3 in right panel of Figure \ref{fig:gamma_maps}). PS3 corresponds to 4FGL-DR3 source J0517.9--6930c. 
A count and TS map between energies 1--10\,GeV are shown in Figure \ref{fig:gamma_maps} where the TS map, right panel, is generated from the global source model, which has no associated source at the position of B0453--685.
Faint $\gamma$-ray emission is visible and coincident with the SNR~B0453--685. 


\begingroup
\begin{table*}
\makebox[0.88\textwidth][c]{
\begin{tabular}{cccccccc}
\hline
\hline
\ Spectral Model & $\log{L}$ & $\Gamma$ & $\alpha$ or $\Gamma_1$ & $\beta$ or $\Gamma_2$ & $G_E$ (MeV cm$^{-2}$ s$^{-1}$) & $E_b$ or $a$ & TS \
\\
\hline
Power law & --505673 & $2.3\pm0.2$ & -- & -- & $ 7.5 (\pm 2.2) \times10^{-7}$ & -- & 23 \\
Log Parabola & --505670 & -- & $2.5 \pm0.4$ & $0.5 \pm0.3$ & $ 5.0 (\pm 1.4) \times10^{-7}$ & 4000 & 27 \\
Power Law with Exponential Cut-Off & --505673 & -- &  $0.8\pm 0.8$ & $0.7$ (fixed) & $5.1 (\pm 1.3) \times10^{-7}$ & $0.009 \pm 0.005$ &  27 \\
\hline
\hline
\end{tabular}}
\caption{Summary of the best-fit parameters and the associated 68\% C.L. statistics for all point source models tested. $G_E$ is the integrated energy flux for energies 300\,MeV--2\,TeV. The units for $E_b$ are MeV. The units for the exponential factor $a$ are MeV$^{-\Gamma_2}$.}
\label{tab:gamma}
\end{table*}
\endgroup

\subsubsection{\fermi Data Analysis Results}\label{sec:fermi-results}

To model the $\gamma$-ray emission coincident with B0453--685 we add a point source at the PWN location (R.A., Dec.) J2000 = (73.408\,$\degree$, --68.489\,$\degree$) to the 300\,MeV--2\,TeV source model. 
With a fixed location, we set the spectrum to a power law characterized by a photon index $\Gamma=2$,
\begin{equation}
  \frac{dN}{dE} = N_{0} \big(\frac{E}{E_0}\big)^{-\Gamma}
\end{equation}
 $E_0$ is set to 1000\,MeV. We then allow the spectral index and normalization to vary. The TS value for a point source with a power law spectrum and photon index, $\Gamma=2.3\pm0.2$, is $23$. We investigate the spectral properties of the $\gamma$-ray emission by changing the spectral model to a log parabola shape following the definition\footnote{For a review of \fermi source spectral models see \url{https://fermi.gsfc.nasa.gov/ssc/data/analysis/scitools/source_models.html}.},
\begin{equation}
  \frac{dN}{dE} = N_{0}\big(\frac{E}{E_b}\big)^{(\alpha+\beta\log{E/E_b})}
\end{equation}
We fix $E_b = 4.0$\,GeV but allow $\alpha$, $\beta$, and $N_0$ to vary in the fit. The TS value of a point source at the PWN/SNR position with a log parabola spectrum is $27$ and has $\alpha=2.5\pm0.4$ and $\beta=0.5\pm0.3$. We test the spectral parameters once more using a spectrum typically observed with MeV--GeV pulsars, a power law with a super-exponential cut-off (PLEC)\footnote{This follows the \texttt{PLSuperExpCutoff2} form used for the 4FGL--DR2. Details can be found here: \url{https://fermi.gsfc.nasa.gov/ssc/data/analysis/scitools/source\_models.html\#PLSuperExpCutoff2}}:
\begin{equation}
  \frac{dN}{dE} = N_{0} \big(\frac{E}{E_0}\big)^{-\Gamma_1} \exp{\big(-a E^{\Gamma_2}\big)}
\end{equation}
 where $E_0$ is the scale (set to 1000\,MeV). 
The TS value of a point source at the position of B0453--685 with a PLEC spectrum is $27$ and has $\Gamma_1=0.8\pm0.8$, $\Gamma_2$ is fixed to $0.7$, and exponential factor $a = 0.009 \pm 0.005$, which corresponds to a $E_c \sim 1\,$GeV energy cut-off. See Table \ref{tab:gamma} for a summary of the spectral parameters for each point source test. 

\begin{figure}
\centering
\includegraphics[width=0.5\textwidth]{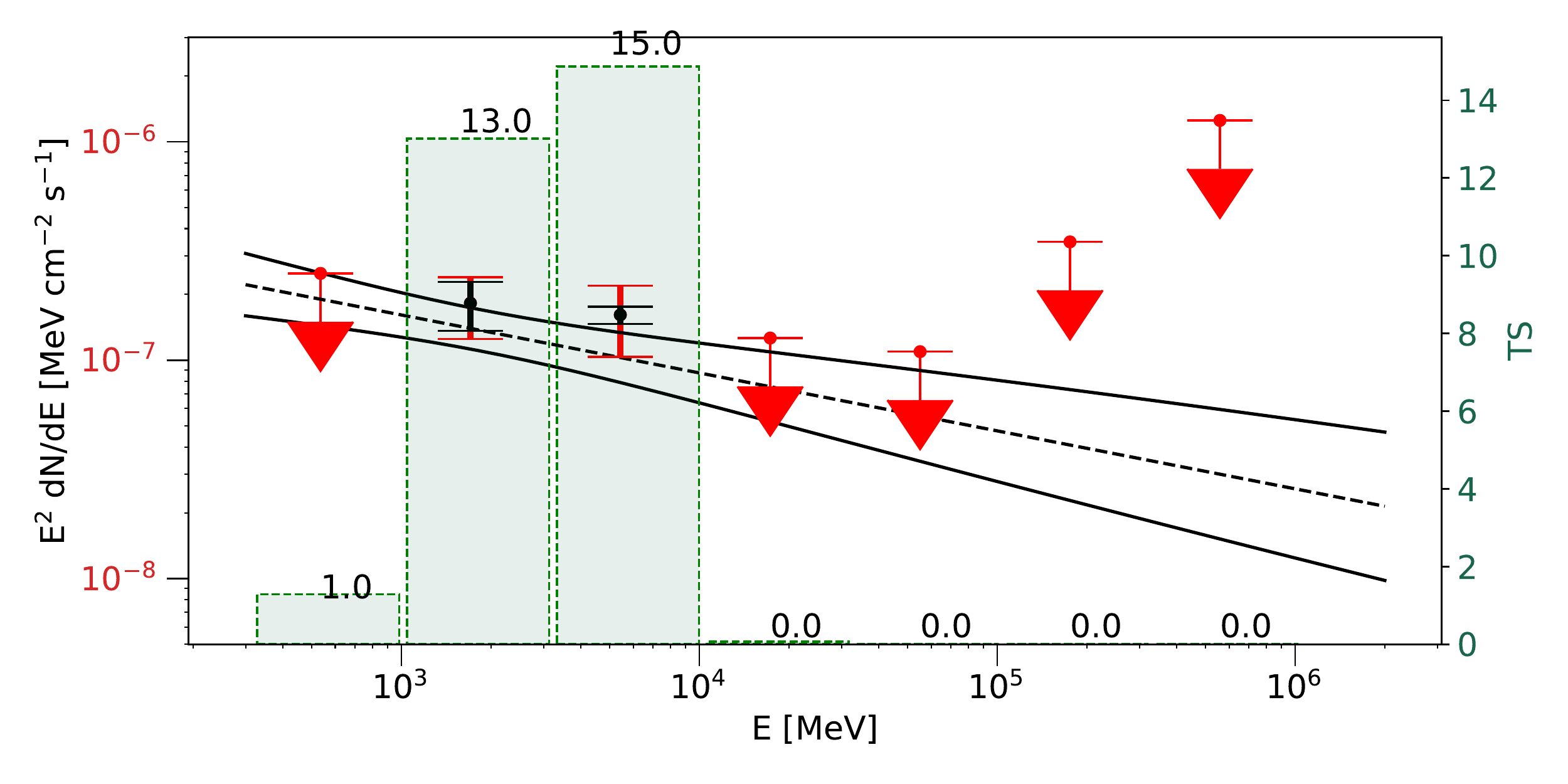}
\caption{The best-fit $\gamma$-ray SED for B0453--685 with 1-$\sigma$ statistical uncertainties in red for $TS > 1$ and 95\% confidence level (C.L.) upper limits otherwise. 
The systematic error from the choice of diffuse LMC model is plotted in black. TS values for each spectral bin are plotted as the green histogram. The data are best characterized as a power-law with $\Gamma = 2.3 \pm 0.2$.}\label{fig:gamma_sed}
\end{figure}

\fermi pulsars are often characterized as either a power-law or a PLEC spectrum and typically cut off at energies $<10$\,GeV \citep[e.g.,][]{2pc}. While we cannot firmly rule out that the observed $\gamma$-ray emission is from the still-undetected pulsar based on the best-fit spectral parameters, it seems unlikely given the majority of the emission is measured in 1--10\,GeV. 
Between the three tested spectral models, the log parabola and PLEC are only marginally preferred (e.g., $TS_{\texttt{LogParabola}} = 2 \Delta(\ln{\mathcal{L}})=5.9$) and carry another degree of freedom with respect to the power law spectral model. We therefore conclude that the best characterization for the $\gamma$-ray emission coincident with SNR~B0453--685 is a power-law spectrum. The corresponding $\gamma$-ray SED is displayed in Figure~\ref{fig:gamma_sed}. 

We localize the point source modeled using a power-law spectrum with \texttt{GTAnalysis.localize} to find the best-fit position and uncertainty. The localized position for the new $\gamma$-ray source is offset by 0.01\,$\degree$ from the exact position of B0453--685 and has R.A., Dec. = 73.39\,$\degree$, --68.49\,$\degree$ (J2000). The corresponding 95\% positional uncertainty radius is $r=0.12 \degree$.
We run extension tests for the best-fit point source in FermiPy utilizing \texttt{GTAnalysis.extension} and the two spatial templates supported in the FermiPy framework, the radial disk and radial Gaussian templates. Both of these extended templates assume a symmetric 2D shape with width parameters radius and sigma, respectively. We fix the position but keep spectral parameters free to vary when finding the best-fit spatial extension for both templates. The best-fit parameters for the extension tests are presented in Table~\ref{tab:extent}. 
The faint $\gamma$-ray source does not display significant extension, consistent with the size of B0453--685 if observed by Fermi. We also perform a variability analysis following the method in the 4FGL catalogs using 1-year time bins. There is no significant variability observed ($TS_{var} < 2$). Finally, we search the new $\gamma$-ray source's 95\% uncertainty region for the spatial overlap with any other objects that may be able to explain the observed $\gamma$-ray emission. There are more than 150 LMC stars within the confidence region, but SNR~B0453--685 is the only non-stellar object\footnote{\url{https://simbad.u-strasbg.fr/simbad/sim-fcoo}}. 

\subsubsection{Systematic Error from Choice of IEM and IRF}

We account for systematic uncertainties introduced by the choice of the interstellar emission model (IEM) and the IRFs, which mainly affect the spectrum of the measured $\gamma$-ray emission. We have followed the prescription developed in \citet{depalma2013,acero2016}, based on generating eight alternative IEMs using a different approach than the standard IEM \citep[see][for details]{acero2016}. For this analysis, we employ the eight alternative IEMs (aIEMs) that were generated for use on Pass 8 data in the {\it Fermi} Galactic Extended Source Catalog \citep[FGES,][]{ackermann2017}. The $\gamma$-ray point source coincident with SNR~B0453--685 is refit with each aIEM to obtain a set of eight values for the spectral flux that we compare to the standard model following equation (5) in \citet{acero2016}. 

We estimate the systematic uncertainties from the effective area\footnote{\url{https://fermi.gsfc.nasa.gov/ssc/data/analysis/LAT_caveats.html}} while enabling energy dispersion as follows: $\pm 3\%$ for $E < 100$\,GeV, $\pm 4.5 \%$ for $E = 175\,$GeV, and $\pm 8\%$ for $E = 556\,$GeV. Since the IEM and IRF systematic errors are taken to be independent, we can evaluate both and perform the quadratic sum for the total systematic error. We find that the systematic errors are negligible for B0453--685 which is not surprising given the location of the Large Magellanic Cloud with respect to the bright diffuse $\gamma$-ray emission along the Galactic plane. 

\subsubsection{Systematic Error from Choice of Diffuse LMC}

We must also account for the systematic error that is introduced by having an additional diffuse background component. This third component is attributed to the cosmic ray (CR) population of the Large Magellanic Cloud interacting with the LMC ISM and there are limitations to the accuracy of the background templates used to model this emission, similar to the Galactic diffuse background. We can probe these limitations by employing a straightforward method described in \citet{lmc2016} to measure systematics from the diffuse LMC. This requires replacing the four extended sources that represent the diffuse LMC in this analysis \citep[the {\it emissivity model},][]{lmc2016} with four different extended sources to represent an alternative template for the diffuse LMC \citep[the {\it analytic model},][]{lmc2016}. The $\gamma$-ray point source coincident with SNR~B0453--685 is then refit with the alternative diffuse LMC template to obtain a new spectral flux that we then compare with the results of the {\it emissivity model} following equation (5) in \citet{acero2016}. The systematic error from the choice of the diffuse LMC template is largest in the two lowest-energy bins, but 
negligible in higher-energy bins. The corresponding systematic error is plotted in Figure~\ref{fig:gamma_sed} in black.

\begingroup
\begin{table}
\centering
\begin{tabular}{cccccc}
\hline
\hline
\ Spatial Template & TS & TS$_{ext}$ & 95\% radius upper limit ($\degree$) \
\\
\hline
Point Source & 23 & -- & -- \\
Radial Disk & 23 & 0.1 & 0.2 \\
Radial Gaussian & 23 & 0.1 & 0.2 \\
\hline
\hline
\end{tabular}
\caption{Summary of the best-fit parameters and the associated statistics for each spatial template used in our analysis.}
\label{tab:extent}
\end{table}
\endgroup





\section{Broadband modeling}

\subsection{Investigating Gamma-ray Origin} \label{sec:naima_model}

For a $\gamma$-ray source at $d = 50\,$kpc, the 300\,MeV--2\,TeV $\gamma$-ray luminosity is $L_\gamma = 2.6 \times 10^{35}$\, erg s$^{-1}$. We compare this value and the best-fit spectral index $\Gamma_\gamma = 2.3$ to Figure~17 in \citet{acero2016} which plots the GeV luminosity against the power-law index for \fermi detected SNRs. There is a correlation between the GeV properties and age of a SNR, in particular the softest (i.e., oldest) SNRs have larger GeV luminosities than harder (i.e, younger) SNRs. Comparing the GeV luminosity found here to those shown in Figure~17, we see that the $\gamma$-ray source is in agreement with the evolved SNRs. This observed correlation is likely due to evolved SNRs interacting with dense material \citep{acero2016}, yet the SNR shell associated to B0453--685 does not show compelling evidence for such an interaction (Figure~\ref{fig:radio_xray_halpha}). We also compare the GeV luminosity to those of \fermi detected pulsars and PWNe \citep{2pc,acero2013}, finding that the GeV luminosity is characteristic of both source classes. Moreover, the spectral index $\Gamma_\gamma = 2.3 \pm 0.2$ is in agreement with \fermi detected SNRs, PWNe, and pulsars: $\Gamma_{\gamma, \text{SNRs}} \approx 2.3$, $\Gamma_{\gamma, \text{PWNe}} \approx 2.1$, and $\Gamma_{\gamma, \text{PSRs}} \approx 2.3$ are the average power-law indices for SNRs, PWNe, and pulsars in the 4FGL-DR2 catalog, respectively \citep[][]{4fgl-dr2}.


\begin{figure*}[htbp]
\begin{minipage}[b]{1.0\textwidth}
\centering
\includegraphics[width=0.75\linewidth]{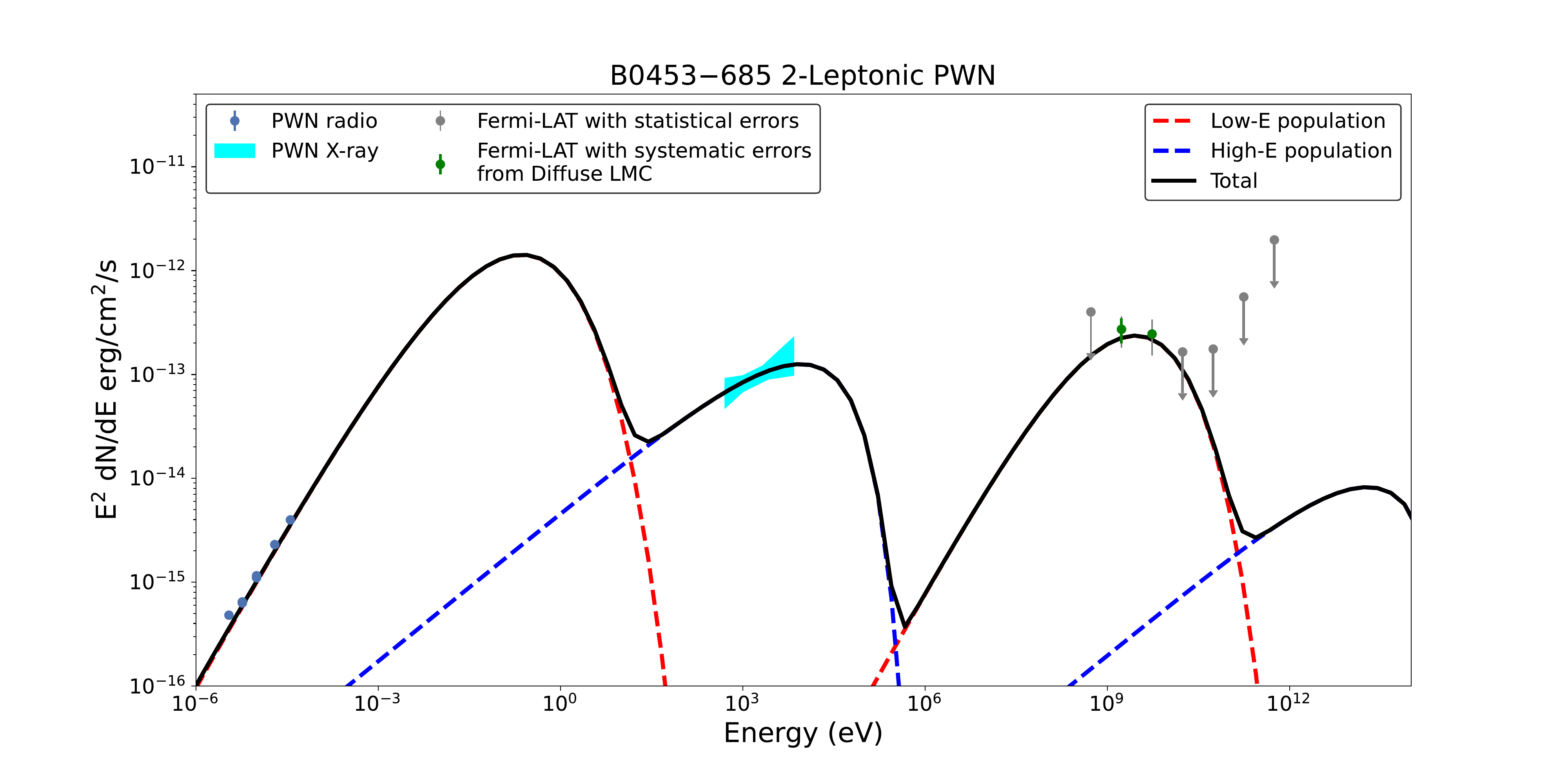}
\vspace{-0.35cm}
\end{minipage}
\begin{minipage}[b]{1.0\textwidth}
\centering
\includegraphics[width=1.0\linewidth]{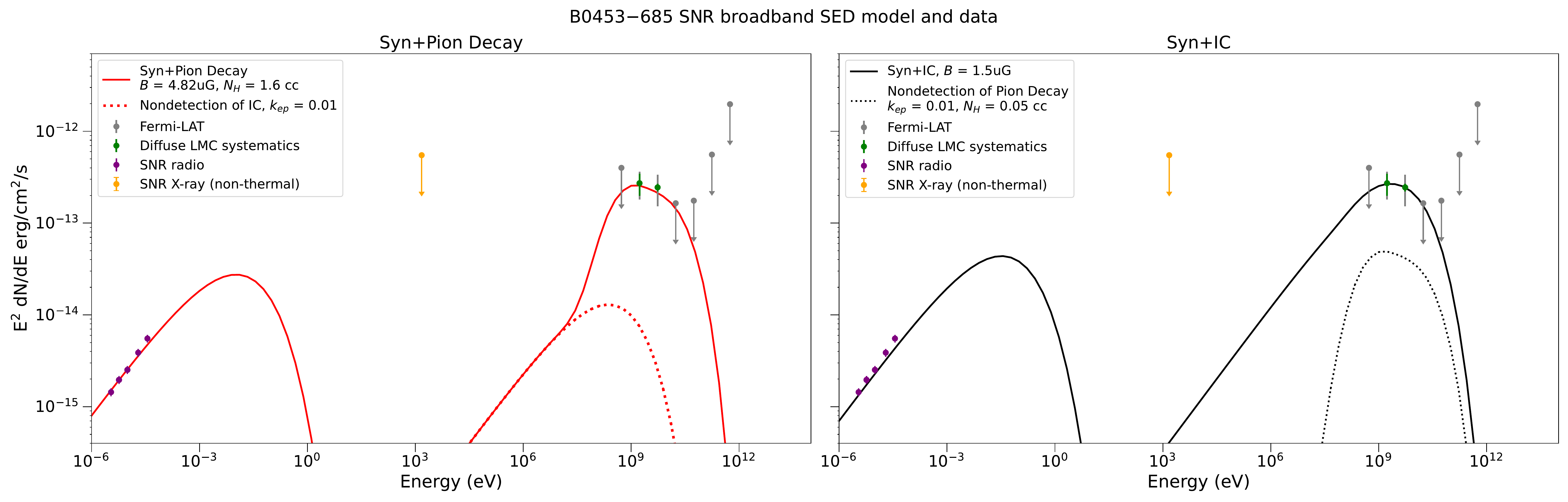}
\end{minipage}
\vspace{-0.5cm}
\caption{The best-fit broadband models for the three scenarios investigated to understand $\gamma$-ray origin. {\it Top:} Two leptonic populations are required to explain the broadband PWN emission. {\it Bottom Left:} a single leptonic population describing SNR synchrotron emission combined with a single hadronic population describing the $\gamma$-ray emission via pion decay from the SNR. {\it Bottom Right:} The case where the leptonic population dominates over the hadronic population in the SNR. 
Radio data of PWN (blue) and SNR (purple) are from \citet{haberl2012}, X-ray data of PWN (cyan) and SNR (yellow) are described in detail in Section \ref{sec:cha-results}, and $\gamma$-ray data (grey/green) in Section \ref{sec:fermi-results}. The uncertainties to the corresponding radio data are very small at this flux scale.}\label{fig:broadbandpwnmodel}
\end{figure*} 

In order to investigate the origin of the observed $\gamma$-ray emission, we use the NAIMA Python package \citep[v0.10.0][]{naima}, which computes the radiation from a single non-thermal relativistic particle population and performs a Markov Chain Monte Carlo (MCMC) sampling of the likelihood distributions \citep[using the \texttt{emcee} package,][]{emcee}. 
For the particle distribution in energy, we assume a power law shape with an exponential cut-off,
\begin{equation}
f(E) = A \bigg(\frac{E}{E_0}\bigg)^{-\Gamma} \exp\big({-{\frac{E}{E_{c}}}}\big) \\
\end{equation}
We then test a combination of free parameters (namely the normalization $A$, index $\Gamma$, energy cut-off $E_{c}$, and magnetic field $B$) that can best explain the broadband spectra for the SNR and PWN independently. We consider only one photon field in all Inverse Compton Scattering calculations in this section, the Cosmic Microwave Background (CMB).

\begin{table*}[tbh!]
\centering
\begin{tabular}{| c c c c |}
\hline
\ & Two-Leptonic PWN & Hadronic-dominant SNR & Leptonic-dominant SNR \\
\ & Population~1 \, Population~2 & (Hadrons only) & (Leptons Only) \\
\hline 
Maximum Log Likelihood (DOF) & --2.07 (13--4)\, --8.71 (13--3) & --2.06 (13--3) & --1.67 (13--4) \\
\hline
\hline
\ Maximum Likelihood values & & &\\
\hline
\  $W_e$ or $W_p^{a}$ & $2.84 \times 10^{49}$ \, $1.43 \times 10^{47}$ & $3.89 \times 10^{51}$ & $2.71 \times 10^{50}$ \\
\ Index & 0.88 $\pm\, 0.13$ \, 2.05 $\pm\, 0.62$ & 1.95 (fixed) & 1.95 $\pm\, 0.05$ \\
\ $\log_{10}{E_{c}}^{b}$ & --0.45 $\pm\, 0.11$ \, 2.35 $\pm\, 0.26$ & --0.68 $\pm\, 0.41$ & --0.17 $\pm\, 0.15$\\
\ $B^{c}$ & 8.18 $\pm\, 4.25$ \, \, 8.18 \text(fixed) & 4.82 $\pm\, 0.12$ & 1.47$\pm\, 0.29$ \\
\hline
\end{tabular}
\caption{Summary of the statistics and best-fit models for the PWN and SNR broadband models displayed in Figure~\ref{fig:broadbandpwnmodel}. The maximum log likelihood can be understood as $\chi^2 = - 2 \ln{L}$. }\footnotesize{$^a$ The total particle energy $W_e$ or $W_p$ is in unit ergs, $^b$ Logarithm base 10 of the cutoff energy in units TeV, $^c$ magnetic field in units $\mu$Gauss}
\label{tab:naima}
\end{table*}

\subsubsection{PWN as Gamma-ray origin}
The radio spectrum considered together with the hard X-ray spectrum of the PWN strongly indicate the presence of more than one particle population, which is also indicated by the estimated age and evolutionary phase of the host SNR. Moreover, the observed X-ray morphology of the nebula displays features consistent with an evolved SNR where the reverse shock has impacted the PWN, compressing the population of previously injected particles while the central pulsar continues to inject new high-energy particles \citep[e.g.,][]{gaensler2003,haberl2012}. The return of the reverse shock would additionally explain the significantly enhanced abundances relative to the local ISM, indicating the PWN plasma is becoming ejecta-dominated \citep{mcentaffer2012}. Based on this, we instead incorporate two leptonic particle populations under the same conditions (nebular magnetic field and ambient photon fields) and combine them to represent a two-leptonic broadband model. A two-leptonic broadband model can describe well the PWN radio, X-ray, and $\gamma$-ray data, where the lower-energy particles dominate the radio and $\gamma$-ray emission while the higher-energy particles are losing more energy in synchrotron radiation than in IC radiation, and therefore dominate in X-ray. We allow Population~1, the lower-energy population, to constrain the magnetic field strength, as the oldest particles likely dominate the synchrotron emission \citep{gelfand_2009}. It is possible each population is interacting with magnetic field regions of varying strength, but for simplicity, we fix the magnetic field value to the best-fit found from the lower-energy population's broadband model when searching for a model fit for the higher-energy population, $B \sim 8 \mu$G. The best-fit parameters for the low-energy population are $\Gamma = 0.88 \pm 0.13$ and $E_{c} = 0.35 \pm 0.11$\,TeV. The best-fit parameters for the high-energy population are $\Gamma = 2.05 \pm 0.62$ and $E_{c} = 224_{-101}^{+183}$\,TeV. The best-fit two-leptonic broadband model for the PWN is displayed in the top panel of Figure~\ref{fig:broadbandpwnmodel} and the corresponding best-fit parameters for both particle populations are listed in Table \ref{tab:naima}. 

The two-leptonic broadband model for the PWN has an estimated total particle energy $W_e = 2.86 \times 10 ^{49}$\,erg. The lower-energy population is responsible for $W_e = 2.84 \times 10 ^{49}$\,erg and the higher-energy population with the remainder, $W_e = 1.43 \times 10 ^{47}$\,erg. 


\subsubsection{SNR as Gamma-ray origin}
There are two possible scenarios for the SNR to be responsible for the $\gamma$-ray emission. The first is a single leptonic population that is accelerated at the SNR shock front, generating both synchrotron emission at lower energies and IC emission at higher energies in $\gamma$-rays \citep[e.g.,][]{reynolds_2008}. The second scenario is a single leptonic population emitting mostly synchrotron radiation at lower energies combined with a single hadronic population emitting $\gamma$-rays through pion decay. 
We describe both of these models and their implications below. 

To model the lower energy SNR emission together with the newly discovered \fermi emission using a single leptonic population (i.e., the leptonic-dominant scenario), we require a particle index $\Gamma = 1.95 \pm 0.05$, an energy cut-off at 671\,GeV, and an inferred magnetic field $B = 1.47\,\mu$G. For the hadronic-dominant scenario, we model the broadband SNR emission assuming a single leptonic and single hadronic population. We measure the magnetic field value to be $B = 4.82\pm\,0.12\,\mu$G for the synchrotron component under the electron-to-proton ratio assumption $k_{ep} = 0.01$ \citep{castro_2013} and characterizing the $\gamma$-ray emission via pion decay through proton-proton collisions at the SNR shock front. The pre-shock proton density $n_0$ has been estimated to be $\sim 0.4$\,cm$^{-3}$ from the SNR X-ray emission measured along the rim region \citep{gaensler2003}. The post-shock proton density at the SNR forward shock $n_H$ could be about four times as high as $n_0$; thus for a compression ratio $\frac{n_H}{n_0} = 4$, $n_H \sim 1.6\,$cm$^{-3}$ \citep[e.g.,][]{vink2003}. We fix the target proton density $n_H = 1.6\,$cm$^{-3}$ at the default differential cross-section \citep[\texttt{Pythia8},][]{naima} while also fixing the proton particle index to $\Gamma = 1.95$. The latter choice is motivated by the particle index being well-defined from the radio data in the leptonic population, but is not well constrained for the hadronic population. We measure an energy cut-off $E_c = 0.194_{-0.11}^{+0.27}$\,TeV for the proton spectrum that can best reproduce the observed $\gamma$-ray spectrum. 
The best-fit broadband models for the SNR are displayed in the lower panels of Figure~\ref{fig:broadbandpwnmodel} and the corresponding parameters are listed in Table \ref{tab:naima}. 


The best-fit leptonic-dominant model for the SNR yields a total electron energy $W_e = 2.71 \times 10^{50}$\,erg. This implies, assuming $k_{ep} = 0.01$, the total proton energy from undetected pion decay emission is $W_p = W_e \times 100 = 2.71 \times 10^{52}$\,erg, requiring roughly 20 times the canonical expectation $E = 10^{51}$\, ergs be in total SNR CR energy alone and a very low target density $n_H = 0.05$\,cm$^{-3}$. 
The best-fit hadronic-dominant model requires a total proton energy $W_p = 3.89\times 10 ^{51}$\,erg, a factor of almost 4 times greater than the typical supernova explosion energy $E = 10^{51}$\, ergs. 

Furthermore, the inferred magnetic field $B = 1.47\,\mu$G in the leptonic-dominant model is comparable to the coherent component of the LMC magnetic field $B \sim 1\,\mu$G \citep{gaensler2005}, which is weaker than one would expect at the SNR shock front, where shock compression can amplify the magnetic field strength 4--5 times the initial value  \citep[see e.g.,][and references therein]{castro_2013}. 
In order to explain the observed $\gamma$-ray emission via pion decay with a reasonable energy in accelerated protons ($E \sim 10^{50}$\,ergs),
the SNR must be interacting with dense ambient material \citep[e.g., similar to W44 and IC443,][]{ackermann2013,chen2014,slane2015}. The radio and X-ray observations of the SNR show a fainter, limb-brightened shell compared to the bright, compact central PWN, providing little evidence of the SNR forward shock colliding with ambient media. 

In conclusion, the energetics inferred from the SNR models lead us to favor the two-leptonic PWN broadband model as the most likely explanation for the $\gamma$-ray emission reported here. We explore the most accurate representation of the PWN broadband data while also exploring the likelihood of a pulsar contribution in the following section. 




\subsection{PWN Evolution through Semi-Analytic Modeling}\label{sec:yosi_model}

\begin{figure}
\hspace{-0.35cm}
\includegraphics[width=0.5\textwidth]{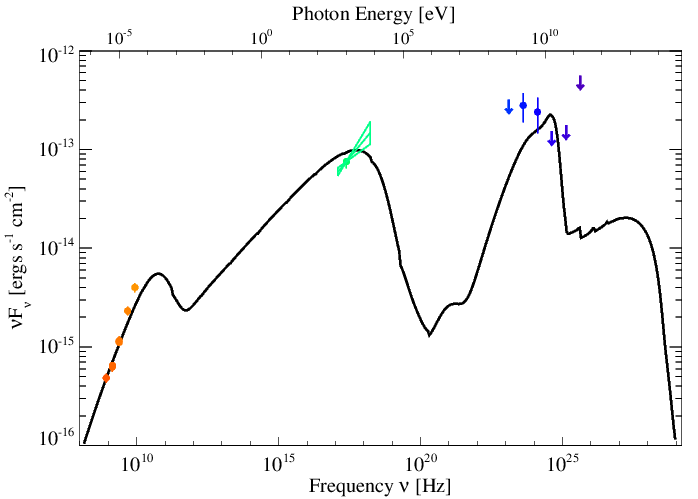}
\caption{The best-fit SED assuming all \fermi emission is non-magnetospheric in origin (i.e., PWN only) obtained through the evolutionary model method described in Section~\ref{sec:yosi_model}. The colored points represent the values of observed data that the model used as comparison points for fitting and are the same values as those in the top panel of Figure~\ref{fig:broadbandpwnmodel}. The small discontinuities in the SED between $\nu \sim 10^{25}-10^{26}\,$Hz are artifacts resulting from the specific numerical implementation of this code and are not physical.}\label{fig:gelfand_sed}
\end{figure} 

We have established in the previous section that modeling the non-thermal broadband SED suggests that it most likely originates from two populations of leptons with different energy spectra, similar to what is expected for evolved PWNe once they have collided with the SNR reverse shock \citep[see e.g.,][]{gelfand_2009, temim2015}. To determine if this depicted scenario can explain the intrinsic properties of this system, we model the observed properties of the PWN, assuming it is responsible for the detected \fermi $\gamma$-ray emission, as it evolves inside the composite SNR B0453$-$685.

\begin{table*}
\centering
\begin{tabular}{|c c c c c|}
\hline
\/Shorthand & Parameter & PWN Best-Fit & PWN$+$PSR Best-Fit & Units \\
\hline
\hline
\ \texttt{loglh} & Log Likelihood of Spectral Energy Distribution & --19.9 & --17.6 & -- \\
\hline
\ \texttt{esn} & Initial Kinetic Energy of Supernova Ejecta & 5.24 & 5.21 & $10^{50}$ ergs \\
\hline
\ \texttt{mej} & Mass of Supernova Ejecta & 2.24 & 2.42 & Solar Masses \\
\hline
\ \texttt{nism} & Number Density of Surrounding ISM & 0.97 & 1.00 & cm$^{-3}$ \\
\hline
\ \texttt{brakind} & Pulsar Braking Index & 2.89 & 2.83 & - \\
\hline
\ \texttt{tau} & Pulsar Spin-down Timescale & 172 & 166 & years \\
\hline
\ \texttt{age} & Age of System & 13900 & 14300 & years \\
\hline
\ \texttt{e0} & Initial Spin-down Luminosity of Pulsar & 6.95 & 6.79 & $10^{39}$ ergs s$^{-1}$ \\
\hline
\ \texttt{etag} & Fraction of Spin-down Luminosity lost as Radiation & $\equiv0$ & 0.246 & - \\
\hline
\ \texttt{etab} & Magnetization of the Pulsar Wind & 0.0006 & 0.0007 & - \\
\hline
\ \texttt{emin} & Minimum Particle Energy in Pulsar Wind & 1.77 & 2.26 & GeV \\
\hline
\ \texttt{emax} & Maximum Particle Energy in Pulsar Wind & 0.90 & 0.73 & PeV \\
\hline
\ \texttt{ebreak} & Break Energy in Pulsar Wind & 76 & 72 & GeV \\
\hline
\ \texttt{p1} & Injection Index below the Break & 1.47 & 1.34 & - \\
\ & (${dN}/{dE} \sim E^{-p1}$) & & & \\
\hline
\ \texttt{p2} & Injection Index below the Break & 2.36 & 2.36 & - \\
\ & (${dN}/{dE} \sim E^{-p2}$) & & & \\
\hline
\ \texttt{ictemp} & Temperature of each Background Photon Field & 1.02 & 1.13 & $10^{6}$ K \\
\hline
\ \texttt{icnorm} & Log Normalization of each Background Photon Field & -17.9 & -18.0 & - \\
\hline
\ \texttt{gpsr} & Photon Index of the $\gamma$-rays Produced Directly by the Pulsar & $\cdots$ & 2.00 & -- \\
\hline
\ \texttt{ecut} & Cutoff Energy from the Power Law of Pulsar Contribution & $\cdots$ & 3.21 & GeV \\
\hline
\end{tabular}
\caption{Summary of the input parameters for the evolutionary system and their best fit values considering PWN-only and PWN+PSR contributions to the \fermi emission.}\label{tab:inputoutputgelfand}
\end{table*}

We use the dynamical and radiative properties of a PWN predicted by an evolutionary model, similar to what is described by \citet{gelfand_2009}, to identify the combination of neutron star, pulsar wind, supernova explosion, and ISM properties that can best reproduce what is observed. The model is developed using a Markov Chain Monte Carlo (MCMC) fitting procedure \citep[see, e.g.,][for details]{gelfand2015} to find the combination of free parameters that can best represent the observations. The observed sizes of the SNR and PWN together with the radio, X-ray and $\gamma$-ray data are used to calculate the final broadband model at an age, $t_{age}$.  The predicted dynamical and radiative properties of the PWN that correspond to the best representation of the broadband data are listed in Table \ref{tab:inputoutputgelfand}. The parameters \texttt{velpsr, etag, kpsr, gpsr,} and \texttt{ecut} are fixed to zero.

The analysis performed here is similar to what has previously been reported for MSH 15--56 \citep{temim2013}, G54.1+0.3 \citep{gelfand2015} G21.5--0.9 \citep{hattori2020}, Kes~75 \citep{gotthelf2021}, and HESS~J1640--465 \citep{mares2021}. For the characteristic age $t_{ch}$ of a pulsar \citep[see][]{pacini1973,gaensler2006}, the age $t_{age}$ is defined as
\begin{equation}
  t_{age} = \frac{2t_{ch}}{p-1} - \tau_{sd}
\end{equation}
and the spin-down luminosity $\dot{E}$ is defined as
\begin{equation}
  \dot{E}(t) = \dot{E_0}\big(1 + \frac{t}{\tau_{sd}}\big)^{-\frac{p+1}{p-1}}
\end{equation}
and are chosen for a braking index $p$, initial spin-down luminosity $\dot{E_0}$, and spin-down timescale $\tau_{sd}$ to best reproduce the pulsar's likely characteristic age and current spin-down luminosity. A fraction $\eta_\gamma$ of this luminosity is converted to $\gamma$-ray emission from the neutron star's magnetosphere, the rest $(1-\eta_\gamma)$ is injected into the PWN in the form of a magnetized, highly relativistic outflow, i.e., the pulsar wind. The pulsar wind enters the PWN at the termination shock, where the rate of magnetic energy $\dot{E}_B$ and particle energy $\dot{E}_P$ injected into the PWN is expressed as:
\begin{eqnarray}\label{eqn:edotb}
\dot{E}_B(t) & \equiv & (1-\eta_\gamma)\eta_{\rm B}\dot{E}(t) \\
\dot{E}_P(t) & \equiv & (1-\eta_\gamma)\eta_{\rm P}\dot{E}(t)
\end{eqnarray}
where $\eta_B$ is the magnetization of the wind and defined to be the fraction of the pulsar's spin-down luminosity injected into the PWN as magnetic fields and $\eta_P$ is the fraction of spin-down luminosity injected into the PWN as particles. We assume the PWN IC emission results from leptons scattering off of the CMB similar to the previous modeling section, however the total particle energy and the properties of the background photon fields cannot be independently determined. Since the evolutionary model accounts for the decline in total particle energy from the adiabatic losses of early PWN evolution and the increase of synchrotron losses at later times from compression, where both likely have a significant effect on the oldest particles, a second photon field is hence required. The second, ambient photon field is defined by temperature $T_{IC}$ and normalization $K_{IC}$, such that the energy density of the photon field $u_{IC}$ is
\begin{equation}
  u_{IC} = K_{IC}a_{BB}T^4_{IC}
\end{equation}
where $a_{BB} = 7.5657 \times 10^{-15}$\,erg cm$^{-3}$ K$^{-4}$. Additionally, we assume the particle injection spectrum at the termination shock is well-described by a broken power law distribution:
\begin{equation}
\frac{d\dot{N}_{e^\pm}(E)}{dE} =
\begin{cases}
 \dot{N}_{break} \big(\frac{E}{E_{break}}\big)^{-p_1} & E_{min} < E < E_{break} \\
 \dot{N}_{break} \big(\frac{E}{E_{break}}\big)^{-p_2} & E_{break} < E < E_{max} \\
\end{cases}
\end{equation}
where $\dot{N}_{e^\pm}$ is the rate that electrons and positrons are injected into the PWN, and $\dot{N}_{break}$ is calculated using
\begin{equation}
  (1-\eta_B)\dot{E} = \int_{E_{min}}^{E_{max}} E \frac{d\dot{N}(E)}{dE} dE
\end{equation}
We show the spectral energy distribution for PWN~B0453--685 that can reasonably reproduce the observed spectrum in Figure~\ref{fig:gelfand_sed}. 

\begin{figure*}
\begin{minipage}[b]{.5\textwidth}
\centering
\includegraphics[width=1.0\textwidth]{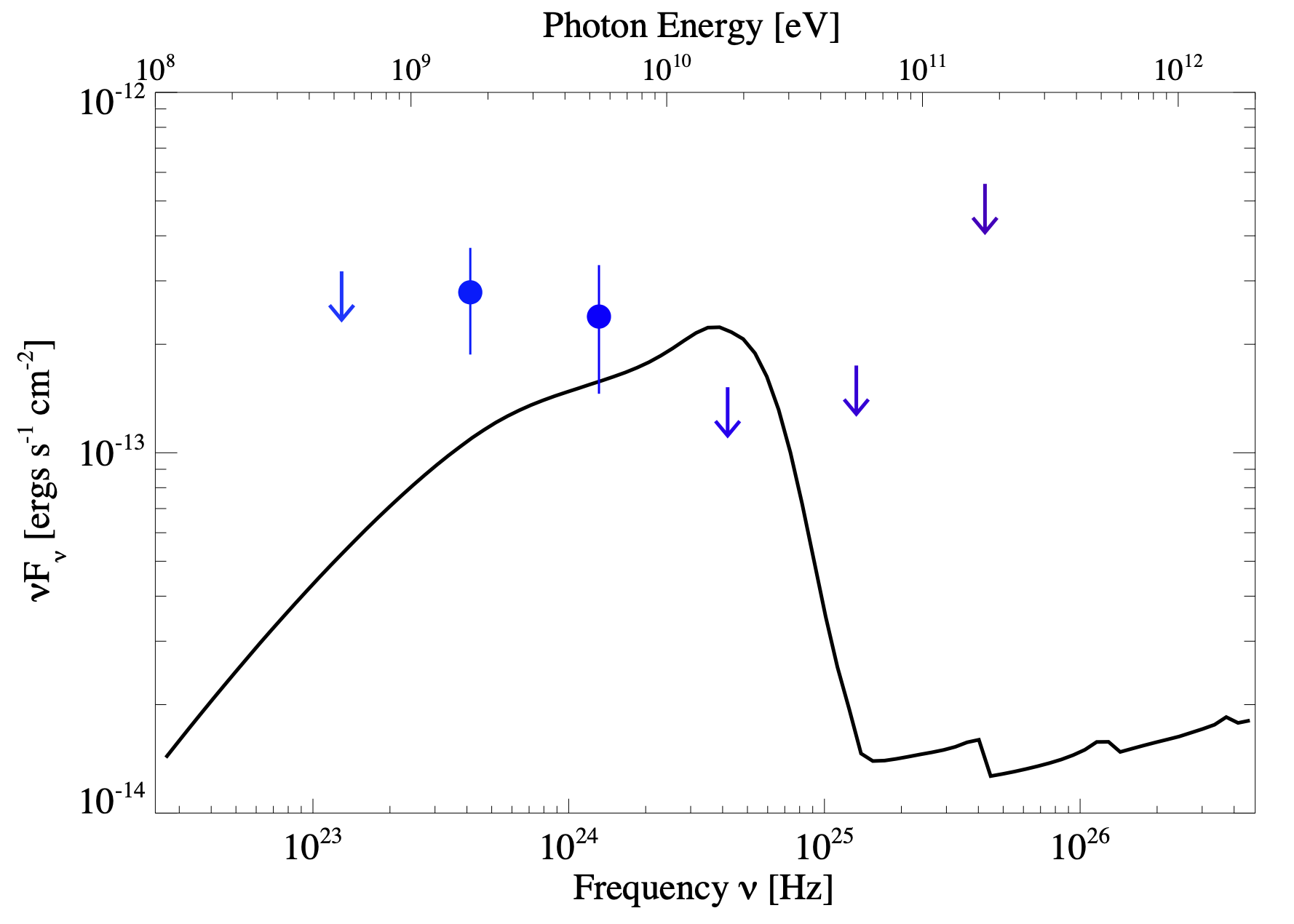}
\end{minipage}
\begin{minipage}[b]{.5\textwidth}
\centering
\includegraphics[width=1.0\textwidth]{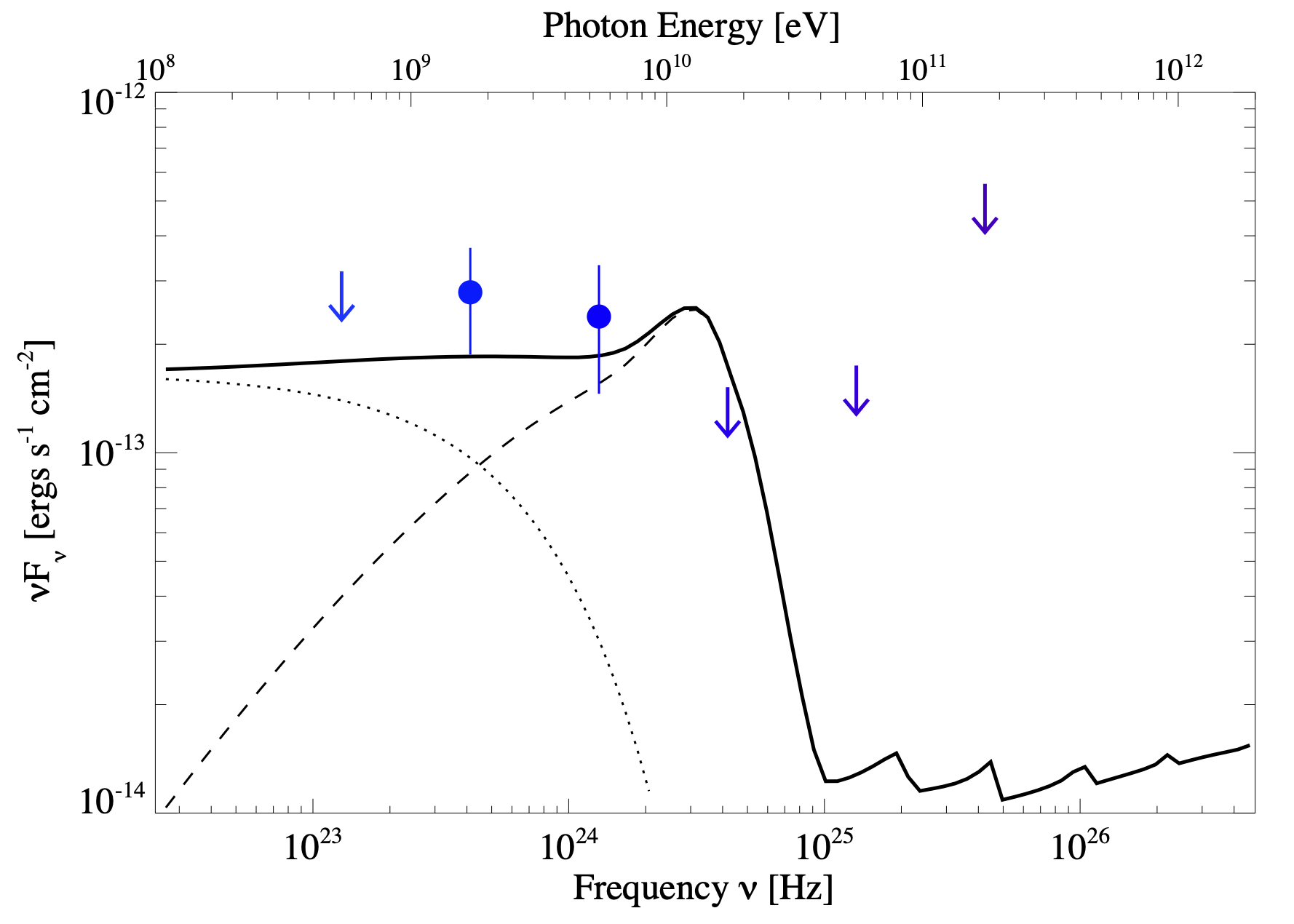}
\end{minipage}
\caption{{\it Left:} The $\gamma$-ray spectral evolutionary model assuming all \fermi emission is non-magnetospheric in origin (i.e., PWN only). {\it Right:} The $\gamma$-ray spectral evolutionary model assuming magnetospheric contribution to the \fermi emission. The dotted line indicates the pulsar contribution and the dashed line indicates the PWN contribution. The colored points represent the values of observed data that the model used as comparison points for fitting and are the same values as those in the top panel of Figure~\ref{fig:broadbandpwnmodel}. In both panels, the discontinuous spectral features beyond $\nu \sim 10^{25}\,$Hz are numerical artifacts and can be ignored.}\label{fig:compare-gelfand-seds}
\end{figure*} 

To investigate the potential for a pulsar contribution to the \fermi data, we model the broadband spectrum again by adding a second emission component from the pulsar. Only the parameter \texttt{velpsr} is fixed to zero. In this case, we assume any \fermi pulsar flux can be described by a power-law with an exponential cut off:
\begin{equation}
  \frac{dN_{\gamma}}{dE} = N_0 E^{-\Gamma} \text{exp}\left(-\frac{E}{E_{\text{cut}}}\right)
\end{equation}
which is a common spectral characterization observed from $\gamma$-ray pulsars \citep{2pc}. We find that the pulsar together with its nebula can readily explain the lower-energy \fermi emission with a cut-off energy $E_c = 3.21\,$GeV and spectral index $\Gamma = 2.0$. The results are similar to the model presented for PWN Kes~75 and its central pulsar \citep{straal2022}. Figure~\ref{fig:compare-gelfand-seds} displays both $\gamma$-ray SEDs for the two considered cases where the \fermi emission is PWN-only (left panel) and where there is both a PWN and pulsar contribution (right panel). If there is a pulsar contribution to the \fermi emission, it is likely to dominate for $E \lesssim 3\,$GeV whereas the PWN may only begin to dominate beyond this energy. We discuss the physical implications of the presented broadband models in the next section. 




\section{Discussion}\label{sec:discuss}
The discovery of faint point-like $\gamma$-ray emission coincident with the SNR~B0453--685 is presented. We can determine the physical properties of the host SNR and ambient medium from the broadband models presented in Sections~\ref{sec:naima_model} and \ref{sec:yosi_model} and compare to the theoretical values expected for a middle-aged SNR in the Sedov-Taylor phase. 

First, we can estimate the post-shock electron density assuming $\frac{n_e}{n_H} = 1.2$ and taking $n_H \sim 1.6\,$cm$^{-3}$ to find $n_e \sim 1.9\,$cm$^{-3}$. This result is consistent with prior works finding a range of values for a filling factor $f$, $n_e/f \sim 1.5 - 8.0\,$cm$^{-3}$ \citep[where a uniform density has $f=1$,][]{gaensler2003,haberl2012,mcentaffer2012}. The post-shock proton density $n_H = 1.6\,$cm$^{-3}$ is less than the average pre-shock LMC ISM density $n_0 \sim 2.0\,$cm$^{-3}$ \citep{kim2003}. The total proton energy and the post-shock proton density characterizing pion decay emission are inversely proportional. If we assume $n_H$ is the expected shock-compressed LMC ISM density then $n_H = 8.0\,$cm$^{-3}$. This would scale down the total energy in protons by a factor $\frac{n_{h, LMC}}{n_{h, \text{X-ray}}} = 5$ to $W_p \sim 7.8 \times 10^{50}\,$erg. This is a more reasonable particle energy, but both SNR models challenge the X-ray observations of the SNR shell, which indicate an explosion energy as low as $E_{SN} \approx 10^{50}$\,erg \citep{gaensler2003,haberl2012}.


The angular diameter of SNR~B0453--685 in both radio and X-ray is 0.036\,$\degree$ (Figure~\ref{fig:radio_xray_halpha}) which corresponds to a shock radius $R_s = 15.7$\,pc at a distance $d = 50$\,kpc. We can evaluate the SNR age assuming it is in the Sedov-Taylor phase \citep{sedov1959}:
\begin{equation}
  \tau = \bigg(\frac{R_{s}}{2.3\text{\,pc}} \big(\frac{E}{10^{51}\text{\,ergs}}\big)^{\frac{1}{5}}\big(\frac{\rho_{0}}{10^{-24}\text{\,g cm$^{-3}$}}\big)^{-\frac{1}{5}}\bigg)^{5/2} 100\, \text{yr}
\end{equation}
The SNR age estimates range between 13\,kyr \citep{gaensler2003} using $E = 5 \times 10^{50}$\,erg and $\rho_0 = m_H n_0 = 0.4 \times 10^{-24}$\,g cm$^{-3}$ where $m_H$ is the mass of a H atom, and 15.2\,kyr using $E = 7.6 \times 10^{50}$\,erg and $\rho_0 = 0.3 \times 10^{-24}$\,g cm$^{-3}$ \citep{haberl2012}. \citet{mcentaffer2012} find the largest age estimates $\tau \sim 17-23$\,kyr using equilibrium shock velocity estimates $\sim 280-380\,$km s$^{-1}$. We adopt the SNR age reported in \citet{gaensler2003} , $\tau \sim 13$\,kyr, which corresponds to a shock velocity $v_s = 478\,$km s$^{-1}$ from $v_s = \frac{2R_s}{5t}$. The age predicted from the evolutionary method in Section~\ref{sec:yosi_model}, $t \sim 14.3$\,kyr, is in good agreement with prior work. The ambient proton density predicted in Section~\ref{sec:yosi_model}, $n_0 = 1.0$\, cm$^{-3}$, is somewhat higher than the values estimated in prior work \citep{gaensler2003,haberl2012}. In any case, the $n_0$ estimates are much lower than the average LMC ISM value $n_0 \sim 2$\, cm$^{-3}$ \citep{kim2003}, and indicate that the ambient medium surrounding SNR~B0453--685 may be less dense than the average LMC ISM. This is supported by Figure~\ref{fig:radio_xray_halpha}, left panel, where a possible density gradient decreasing from east to west is apparent. While H$\alpha$ emission is not a direct tracer for molecular material, it is a byproduct of SNRs interacting with molecular material \citep[e.g.,][]{winkler2014,eagle2019}. The lower ambient particle density estimate is also consistent with the observed faint SNR shell in radio and X-ray. It therefore seems unlikely for the SNR to be the gamma-ray origin, whether leptonic or hadronic.

We instead favor a model where the observed $\gamma$-rays are produced by an energetic neutron star and its resultant PWN, which can adequately describe the observed properties of this system as detailed in Section \ref{sec:yosi_model}.
The explosion energy predicted by the evolutionary model, $E = 5.2 \times 10^{50}$\,erg, is very similar to that inferred by X-ray observations, $E \sim 5-7.6 \times 10^{50}$\,erg \citep{gaensler2003,haberl2012}. Additionally, the magnetic field and total particle energy in the PWN from the evolutionary model are predicted to be $5.9\,\mu$G and $W_e = 5.4 \times 10^{48}\,$erg respectively, which is roughly consistent to the values implied by NAIMA modeling in Section~\ref{sec:naima_model}, $8.18\,\mu$G and $W_e = 2.9 \times 10^{49}\,$erg. Lastly, one can estimate the $\gamma$-ray efficiency $\eta = \frac{L_\gamma}{\dot{E}}$ from the predicted current spin-down power of the central pulsar in the evolutionary model, $\dot{E} \sim 8.1 \times 10^{35}$\, erg s$^{-1}$. For a 300\,MeV--2\,TeV $\gamma$-ray source at $d = 50\,$kpc, the $\gamma$-ray luminosity is $L_\gamma = 2.6 \times 10^{35}$\, erg s$^{-1}$ which corresponds to $\eta = 0.32$. This efficiency value is not uncommon for $\gamma$-ray pulsars \citep[e.g.,][]{2pc}, though it is a more compatible value to expect from evolved PWNe.

From the presented semi-analytic evolutionary models, we find the best representation of the data occurs with the supernova energy values $\sim 5 \times 10^{50}$\,erg, $\sim 2.3$ solar masses for SN ejecta, and $\sim$1.0 cm$^{-3}$ for the density of the ISM (see Table~\ref{tab:inputoutputgelfand}). These values can then be used in combination with other models to survey the possible physical characteristics of the progenitor for SNR~B0453--685. For example, a correlation reported in \citet{ertl2020} has found that the only supernovae that have an explosion energy $\sim 5 \times 10^{50}$\,erg are those whose progenitors have a final helium core mass $< 3.5\,M_\odot$. Given an ejecta mass $\sim 2.3\,M_\odot$ from the presented evolutionary model, we calculate a neutron star mass $M_{NS} = 3.5\,M_\odot - 2.3\,M_\odot = 1.2\,M_\odot$, which is reasonable \citep[see e.g.,][]{kaper2006}. 

A core collapse supernova progenitor cannot have an initial mass smaller than $8\,M_\odot$. We can use the known inverse correlation between the age and mass of a main-sequence star,
\begin{eqnarray}
\label{eqn:age_mass}
  \frac{t_{MS}}{t_{Sun}} \sim \big(\frac{M}{M_{Sun}}\big)^{-2.5}
\end{eqnarray}
to get a maximum possible lifetime $\tau \sim 20\,$million\,years for any supernova progenitor. A map by \citet{harris2009} of the LMC with age and metallicity data distributions provides the age and metallicity distributions for the LMC regions closest to the location for B0453--685. 
By compiling the data in \citet{harris2009}, we can see that there was possibly a burst of star formation in those regions around the maximum possible lifetime estimate, as it contains many stars that are from 10$^{6.8}$ ($\sim 6.3$\,million) to 10$^{7.4}$ ($\sim 25$\,million) years old. From this, the progenitor would have had a main sequence lifetime comparable to the maximum possible lifetime for us to observe the supernova remnant today. We can use Eq.~\ref{eqn:age_mass} to estimate the mass of the precursor star of B0453--685 to be between 11 and 19\,$M_\odot$. However, as said above, the presented model predicts a pre-explosion helium core of 3.5 solar masses, which does not reach the 11--19\,$M_\odot$ dictated by the above analysis. The similarity between the inferred final core mass $M_{NS} = 1.2\,M_\odot$ suggested by the presented modeling and the predicted pre-explosion mass $M_{\text{pre-explosion}} = 3.5\,M_\odot$ from \citet{ertl2020} implies that the progenitor lost its envelope before exploding.

If the models presented are correct, then there are two plausible ways to explain the loss of $\sim 7.5-15.5\,M_\odot$ of material before exploding: an isolated star could have lost mass by way of stellar wind, while a star that is part of a binary system could have transferred some of its mass to the other star. To account for stellar wind quantitatively, we looked at the model presented in \citet{sukh2016} where it is shown that normal ejecta mass for a 10--15$\,M_\odot$ star is 8--10$\,M_\odot$, respectively. However, stellar wind can only account for up to 3$\,M_\odot$ in mass loss for stars more massive than 15$\,M_\odot$. Additionally, it is known that low metallicity stars experience less mass loss \citep{heger2003}, and the young stars in the LMC region of B0453--685 all have metallicity $\sim$ 0.008\,$Z_\odot$. In summary, it seems plausible that the progenitor for B0453--685 was a part of a binary star system. 

\section{Conclusions}\label{sec:conclude}
We have reported the discovery of faint, point-like $\gamma$-ray emission by the \fermi that is coincident with the composite SNR B0453--685, located within the Large Magellanic Cloud. We provide a detailed multiwavelength analysis that is combined with two different broadband modeling techniques to explore the most likely origin of the observed $\gamma$-ray emission. We compare the observed $\gamma$-ray emission to the physical properties of SNR~B0453-685 to determine that the association is probable. We then compare the physical implications and energetics from the best-fit broadband models to the theoretically expected values for such a system and find that the most plausible origin is the pulsar wind nebula within the middle-aged SNR~B0453--685 and possibly a substantial pulsar contribution to the low-energy $\gamma$-ray emission below $E < 5$\,GeV. Theoretical expectation based on observational constraints and the inferred values from the best-fit models are consistent, despite assumptions about the SNR kinematics and environment in the evolutionary modeling method such as a spherically symmetric expansion into a homogeneous ISM density. The MeV--GeV detection is too faint to attempt a pulsation search and the $\gamma$-ray SED cannot rule out a pulsar component. We attempt to model the $\gamma$-ray emission assuming both PWN and pulsar contributions and the results indicate that any pulsar $\gamma$-ray signal is likely to be prominent below $E \leq 5\,$GeV, if present. 
Further work should explore the $\gamma$-ray data particularly for energies $E < 10\,$GeV to investigate the potential for a pulsar contribution as well as the possibilities for PWN and/or pulsar emission in the MeV band for a future MeV space missions such as COSI\footnote{\url{https://cosi.ssl.berkeley.edu/}} and AMEGO\footnote{\url{https://asd.gsfc.nasa.gov/amego/index.html}}. The IC emission spectra reported here may be even better constrained when combined with TeV data from ground-based Cherenkov telescopes such as H.E.S.S. or the upcoming Cherenkov Telescope Array\footnote{\url{https://www.cta-observatory.org/}}. 

\begin{acknowledgements}
The \fermi Collaboration acknowledges generous ongoing support from a number of agencies and institutes that have supported both the development and the operation of the LAT as well as scientific data analysis. These include the National Aeronautics and Space Administration and the Department of Energy in the United States, the Commissariat \a'a l'Energie Atomique and the Centre National de la Recherche Scientifique / Institut National de Physique Nucl\a'eaire et de Physique des Particules in France, the Agenzia Spaziale Italiana and the Istituto Nazionale di Fisica Nucleare in Italy, the Ministry of Education, Culture, Sports, Science and Technology (MEXT), High Energy Accelerator Research Organization (KEK) and Japan Aerospace Exploration Agency (JAXA) in Japan, and the K. A. Wallenberg Foundation, the Swedish Research Council and the Swedish National Space Board in Sweden.

Additional support for science analysis during the operations phase is gratefully acknowledged from the Istituto Nazionale di Astrofisica in Italy and the Centre National d'\'Etudes Spatiales in France. This work performed in part under DOE Contract DE- AC02-76SF00515. Work at NRL is supported by NASA.
\end{acknowledgements}
{\software{CIAO \citep[v4.12][]{ciao2006}, FermiPy \citep[v.1.0.1][]{fermipy2017}, Fermitools: Fermi Science Tools \citep[v2.0.8][]{fermitools2019}, NAIMA \citep{naima}}}
\\
\\
\\

\bibliographystyle{aa}
\bibliography{aa.bib}

\end{document}